\pdfoutput=1
\documentclass[10pt,final,journal]{IEEEtran}

\usepackage{amsmath,amsthm,amssymb,scrextend}
\usepackage{fancyhdr,float,graphicx,afterpage,placeins}
\usepackage{enumitem,tabularx}
\usepackage{tikz}
\usepackage{epstopdf}
\usepackage{parskip}

\DeclareMathOperator*{\argmin}{arg\,min}

\begin{document}
\title{Sensitivity of BPA SAR Image Formation to Initial Position, Velocity, and Attitude Navigation Errors} 
\author{{Colton~Lindstrom, Dr. Randall Christensen, Dr. Jacob Gunther}%
\thanks{This work was supported in part by Sandia National Laboratories, Albuquerque, NM. Navigation and radar data was provided by the Space Dynamics Laboratory, Logan, UT.}
\thanks{C. Lindstrom is with the Electrical and Computer Engineering Department, Utah State University, Logan, UT 84322 USA (e-mail: coltonlindstrom@gmail.com)}%
\thanks{J. Gunther is with the Electrical and Computer Engineering Department, Utah State University, Logan, UT 84322 USA (e-mail: jake.gunther@usu.edu)}%
\thanks{R. Christensen is with the Electrical and Computer Engineering Department, Utah State University, Logan, UT 84322 USA (e-mail: randall.christensen@usu.edu)}}
\maketitle

\begin{abstract} The Back-Projection Algorithm (BPA) is a time domain matched filtering
technique to form synthetic aperture radar (SAR) images. To produce high quality BPA images,
precise navigation data for the radar platform must be known. Any error in position, velocity, or
attitude results in improperly formed images corrupted by shifting, blurring, and distortion.
This paper develops analytical expressions that characterize the relationship between navigation 
errors and image formation errors. These analytical expressions are verified via simulated image
formation and real data image formation.
\end{abstract}

\section{Introduction}\label{sec:introduction}

\IEEEPARstart{S}{ynthetic} aperture radar (SAR) is a class of radar processing that uses the flight
path of a spacecraft or aircraft, referred to as a radar platform, to create a synthetic imaging aperture. 
Through a collection of matched filters, raw radar data is processed into images. Many efficient 
matched filtering algorithms have been developed that employ the frequency domain, such as the 
range-Doppler algorithm, the chirp scaling algorithm, the omega-K algorithm, and more 
\cite{cumming_wong}. Time domain algorithms also exist, such as the Back-Projection Algorithm (BPA) 
\cite{duersch_analysis_2015}. 

This paper explores the sensitivity of BPA images to navigation errors. This is done first analytically
using the range equation and back-projection equation, which are both defined in Section \ref{sec:background}.
Secondly the analysis is verified by injecting error into a flight trajectory estimate of an aircraft and 
using the corrupted trajectory estimate to form BPA images. This process is performed on both simulated 
and real data.

\subsection{Motivation}
The research in this paper is primarily motivated by the field of GPS denied navigation but may be of 
interest to other fields relating to SAR image quality or image autofocusing. GPS denied navigation is 
a field of research that involves estimating the state of a vehicle in the absence of Global Navigation
Satellite Systems (GNSS) such as GPS.  Typical approaches utilize an inertial navigation system (INS) as
the core sensor, aided by measurements from auxiliary sensors, in the framework of an extended Kalman 
filter. Such auxiliary sensors may include cameras, lidar, radar, etc, \cite{balamurugan}.

When forming a SAR image using back-projection, navigation data and raw radar data are processed to form
an image. Obtaining precise navigation data in an ideal application requires the use of GPS. However, in 
a GPS denied environment, navigation errors may be present, which result in distorted SAR images. This 
research is motivated by the potential of inferring navigation errors from induced image errors during BPA
image formation \cite{christensen_radar-aided_nodate}.  This paper works toward building the foundation
and intuition needed to achieve such a potential.

\subsection{Literature Review}
BPA is more sensitive to navigation errors than other types of SAR image formation techniques. This can 
be inferred from Duersch and Long who explore some of the sensitivities inherent in forming images using
back-projection \cite{duersch_analysis_2015}. This research expands the sensitivity analysis to motion 
errors as seen from a navigation point of view with a more complete navigation state. 

BPA is essentially a matched filter along a hyperbolic curve
within a set of range compressed data. Integrating along a curve requires that each data sample be
precisely selected in correspondence with the current position of the vehicle. Any error in 
navigation data results in integrating data on an incorrect curve with an incorrect phase. Very precise
navigation data is therefore necessary to form accurate BPA images 
\cite{duersch_analysis_2015}. Further details on BPA are discussed in Section \ref{sec:background}.

Errors in navigation data manifest themselves in a SAR image as 
shifts and distortions of a given target. Research performed by Christensen et al explores 
the effects of navigation errors on fully formed SAR images and hypothesizes that navigation errors
can be determined by comparing degraded SAR images to a reference SAR map 
\cite{christensen_radar-aided_nodate}.

Many types of errors can affect the quality of SAR images. As such a comprehensive analysis of image
errors is difficult and requires further investigation. Current efforts in analyzing image errors
include research performed by Bamler \cite{bamler_comparison_1992} and
Farrel et al \cite{farrell_effects_1973}. They explore image errors caused by servo transients, 
quantization, range uncertainty, range-rate uncertainty, and focusing algorithm selection. 
Additionally, Chen explores image errors caused by moving targets in the illuminated scene
\cite{chen_time-frequency_1998}.

In previous literature, navigation errors have been expressed as range displacement, line of 
sight displacement, and forward velocity error. Moreira and Xing et al adjust the SAR pulse 
repetition frequency (PRF) to compensate for forward velocity errors 
\cite{moreira_estimating_1990}, \cite{xing_motion_2009}. Moreira further adjusts phase and range 
delays to compensate for line of sight displacement errors. Velocity errors in particular have been
shown to affect the Doppler parameters of the SAR data, which cause target location errors and 
image degradation in the final image \cite{xing_motion_2009}, \cite{bing_image_2013}. 

\subsection{Contributions}
A comprehensive study of BPA SAR image errors in the context of the full navigation state has not
been performed to date.  The research seeks to fill this void by developing relationships between
image shifts, blurs, and distortions and all components of navigation state, specifically 
position, velocity, and attitude errors.

Section \ref{sec:background} begins by providing necessary background knowledge concerning inertial
navigation and BPA processing. Section \ref{sec:analysis} develops the math necessary to predict 
how navigation errors affect the final SAR image. Sections \ref{sec:simulation} and 
\ref{sec:real data} demonstrate the application of the error analysis to simulated and real data, 
respectively. Section \ref{sec:conclusion} provides concluding discussion and summary.

\section{Background}\label{sec:background}
\subsection{Inertial Navigation}
The purpose of this section is to define an inertial navigation framework
applicable to the short data collection times typical of SAR imagery.
The framework is then used to develop analytical expressions of position
estimation error growth.

Inertial navigation is a large field
with an equally-large body of literature dating back to the 1930's.
An excellent overview of the history and motivating factors behind
the development of this field is provided in \cite{grewal_global_2013}.
The navigation framework developed in this section utilizes concepts
discussed in \cite{grewal_global_2013,farrell_aided_2008,savage_strapdown_2000}.
The developed framework is most directly related to the so-called
``Tangent Frame'' kinematic model \cite{farrell_aided_2008}, with
the assumptions of constant gravity and a non-rotating earth, both
of which are applicable over the short time frame typical of an airborne SAR
data collection. In the development that follows, the truth and navigation
states are defined, with the associated differential equations. Consistent
with an extended Kalman filter framework, the truth state differential
equations are linearized about the estimated navigation state to derive
the differential equations of the estimation errors, or error states.

The truth state vector comprises the true position $\left(\boldsymbol{p}^{n}\right)$,
velocity $\left(\boldsymbol{v}^{n}\right)$, and attitude quaternion
$\left(q_{b}^{n}\right)$ of the vehicle
\begin{equation}
\boldsymbol{x}=\left[\begin{array}{ccc}
\boldsymbol{p}^{n} & 
\boldsymbol{v}^{n} & 
q_{b}^{n}
\end{array}\right]^T\label{eq:cov-veh1}
\end{equation}
where $n$ and $b$ refer to the navigation and body frame, respectively.
The body frame origin is coincident with the navigation center of
the inertial measurement unit (IMU), with the axes aligned and rotating
with the vehicle body axes. Out of convenience for the subsequent
analysis, and without loss of generality, the navigation frame is
defined with the $x$-axis parallel to the velocity of the vehicle,
the $z$-axis in the direction of gravity, and the $y$-axis defined
by the right-hand-rule. The $x$, $y$, and $z$ axes, therefore,
correspond to the along-track, cross-track, and down directions typical
of radar imaging conventions. Consistent with Ferell's definition
of the ``Tangent Frame'' \cite{farrell_aided_2008}, the position
and velocity are defined relative to a fixed origin, whose location
is the position of the vehicle at the beginning of the SAR data collection.
The differential equations of the truth states are defined as follows\footnote{In this work, the quaternion is interpreted as a ``left-handed''
quaternion, and the $\otimes$ operator is the Hamiltonian quaternion
product \cite{zanetti_rotations_2019}.}
\begin{equation}
\left[\begin{array}{c}
\dot{\boldsymbol{p}}^{n}\\
\dot{\boldsymbol{v}}^{n}\\
\dot{q}_{b}^{n}
\end{array}\right]=\left[\begin{array}{c}
\boldsymbol{v}^{n}\\
T_{b}^{n}\boldsymbol{\nu}^{b}+\boldsymbol{g}^{n}\\
\frac{1}{2}q_{b}^{n}\otimes\left[\begin{array}{c}
0\\
\boldsymbol{\omega}^{b}
\end{array}\right]
\end{array}\right]\label{eq:cov-veh2}
\end{equation}
 The strapdown inertial navigation system comprises a three-axis accelerometer and gyro, which provide measurements of specific force $\left(\tilde{\boldsymbol{\nu}}^{b}\right)$ and angular rate $\left(\tilde{\boldsymbol{\omega}}^{b}\right)$ in the body frame, corrupted by noise
\begin{equation}
\left[\begin{array}{c}
\tilde{\boldsymbol{\nu}}^{b}\\
\tilde{\boldsymbol{\omega}}^{b}
\end{array}\right]=\left[\begin{array}{c}
\boldsymbol{\nu}^{b}\\
\boldsymbol{\omega}^{b}
\end{array}\right]+\left[\begin{array}{c}
\boldsymbol{n}_{\nu}\\
\boldsymbol{n}_{\omega}
\end{array}\right]\label{eq:cov-cont-inertial-measurements}
\end{equation}

The navigation states are defined identical to the truth states but are propagated using noisy accelerometer and gyro measurements
\begin{equation}
\left[\begin{array}{c}
\dot{\hat{\boldsymbol{p}}}^{n}\\
\dot{\hat{\boldsymbol{v}}}^{n}\\
\dot{\hat{q}}_{b}^{n}
\end{array}\right]=\left[\begin{array}{c}
\hat{\boldsymbol{v}}^{n}\\
\hat{T}_{b}^{n}\tilde{\boldsymbol{\nu}}^{b}+\boldsymbol{g}^{n}\\
\frac{1}{2}\hat{q}_{b}^{n}\otimes\left[\begin{array}{c}
0\\
\tilde{\boldsymbol{\omega}}^{b}
\end{array}\right]
\end{array}\right]\label{eq:cov-veh7}
\end{equation}

The estimation error, or error state vector, 
\begin{equation}\label{eqn:error state position}
\delta\boldsymbol{x}=\left[\begin{array}{ccc}
\delta\boldsymbol{p}^{n} & 
\delta\boldsymbol{v}^{n} & 
\delta\boldsymbol{\theta}{}^{n}
\end{array}\right]^T
\end{equation}
is defined as the difference between the truth states and the navigation
states. For all but the attitude states, the difference is defined
by a simple subtraction. For the attitude quaternion the difference is defined by 
a quaternion product.
\begin{equation}
\delta\boldsymbol{p}^{n}=\boldsymbol{p}^{n}-\hat{\boldsymbol{p}}^{n}\label{eq:cov-calcEstErrors1}
\end{equation}
\begin{equation}
\delta\boldsymbol{v}^{n}=\boldsymbol{v}^{n}-\hat{\boldsymbol{v}}^{n}\label{eq:cov-calcEstErrors2}
\end{equation}
\begin{equation}
\left[\begin{array}{c}
1\\
-\frac{1}{2}\delta\boldsymbol{\theta}{}^{n}
\end{array}\right]=q_{b}^{n}\otimes\left(\hat{q}_{b}^{n}\right)^{*}\label{eq:cov-calcEstErrors3}
\end{equation}
It is also convenient to define the attitude errors in terms of the true and estimated transformation matrices
\begin{equation}
\left[I-\left(\delta\boldsymbol{\theta}{}^{n}\times\right)\right]=T_{b}^{n}\left(\hat{T}_{b}^{n}\right)^{T}\label{eq:cov-calcEstErrors4}
\end{equation}
Linearization of (\ref{eq:cov-veh2}) about the estimated state results in the error state differential equation 
\begin{equation}
\delta\dot{\boldsymbol{x}}=F\delta\boldsymbol{x}+B\boldsymbol{w}\label{eq:cov-err-state-diffeq}
\end{equation}
where the state dynamics matrix, $F$, and noise coupling
matrix, $B$, are defined respectively as 
\begin{equation}
F=\left[\begin{array}{ccc}
0_{3\times3} & I_{3\times3} & 0_{3\times3}\\
0_{3\times3} & 0_{3\times3} & \left[\hat{T}_{b}^{n}\tilde{\boldsymbol{\nu}}^{b}\right]\times\\
0_{3\times3} & 0_{3\times3} & 0_{3\times3}
\end{array}\right]\label{eq:cov-F}
\end{equation}
\begin{equation}
B=\left[\begin{array}{cc}
0_{3\times3} & 0_{3\times3}\\
-\hat{T}_{b}^{n} & 0_{3\times3}\\
0_{3\times3} & \hat{T}_{b}^{n}
\end{array}\right]\label{eq:cov-B}
\end{equation}
and where the white noise $\left(\boldsymbol{w}\right)$ consists
of the accelerometer and gyro measurement noise 
\begin{equation}
\boldsymbol{w}=\left[\begin{array}{cc}
\boldsymbol{n}_{\nu} & 
\boldsymbol{n}_{\omega}
\end{array}\right]^T
\end{equation}
The focus of this paper is to analyze the sensitivity of the
BPA image to errors in position, velocity, and attitude at the
beginning of the synthetic aperture. The effect of $\boldsymbol{w}$
is therefore ignored, and the analysis is facilitated by determining
the homogenous solution to (\ref{eq:cov-err-state-diffeq})
\begin{equation}
\delta\boldsymbol{x}_{k}=\Phi\left(t_{k},t_{k-1}\right)\delta\boldsymbol{x}_{k-1}
\end{equation}
where $\Phi\left(t_{k},t_{k-1}\right)$ is the state transition matrix
(STM) from the $t_{k-1}$ to $t_{k}$. The STM is defined as the matrix
which satisfies the differential equation and initial condition
\begin{equation}
\dot{\Phi}\left(t_{k+1},t_{k}\right)=F\left(t\right)\Phi\left(t_{k+1},t_{k}\right)
\label{eq:cov-stmdot}
\end{equation}
\begin{equation}
    \Phi\left(t_{k},t_{k}\right)=I_{n\times n}
\end{equation}
For the case of straight-and-level flight, the dynamic coupling matrix
of (\ref{eq:cov-F}) is constant, 
\begin{equation}
F=\left[\begin{array}{ccc}
0_{3\times3} & I_{3\times3} & 0_{3\times3}\\
0_{3\times3} & 0_{3\times3} & \left(\boldsymbol{\nu}^{n}\right)\times\\
0_{3\times3} & 0_{3\times3} & 0_{3\times3}
\end{array}\right]\label{eq:cov-F-straight-level}
\end{equation}
Where the accelerometer measurements are expressed in the $n$ frame as
\begin{equation}
\boldsymbol{\nu}^{n}=\left[\begin{array}{ccc}
0 & 0 & -g\end{array}\right]^{T}
\end{equation}
Since $F$ is constant, the STM is derived using the matrix exponential (\cite{maybeck_stochastic_1994}, page 42)
\begin{equation}
\Phi\left(t_{k+1},t_{k}\right)=\left[\begin{array}{ccc}
I_{3\times3} & I_{3\times3}\Delta t & \left(\boldsymbol{\nu}^{n}\times\right)\frac{\Delta t^{2}}{2!}\\
0_{3\times3} & I_{3\times3} & \left(\boldsymbol{\nu}^{n}\right)\times\Delta t\\
0_{3\times3} & 0_{3\times3} & I_{3\times3}
\end{array}\right]
\end{equation}

The desired analytical expression for position errors is obtained
from the first row of the STM to yield
\begin{equation}\label{eqn:position error equation}
\delta\boldsymbol{p}^{n}\left(t\right)=\delta\boldsymbol{p}_{0}^{n}+\delta\boldsymbol{v}_{0}^{n}\Delta t+\boldsymbol{\nu}^{n}\times\delta\boldsymbol{\theta}{}_{0}^{n}\frac{\Delta t^{2}}{2}
\end{equation}


In all subsequent sections, the variables representing positions, velocities, and attitudes are 
all assumed to be in the $n$ frame. As such, the $n$ superscript on all navigation states is 
omitted for notational brevity.

\subsection{Back-Projection Algorithm}
Forming images using SAR is a process of matched filtering that transformed raw returned radar
signals into focused pixels. A raw SAR signal is typically a linear frequency modulated (FM), or
``chirp'', signal. Chirp signals are a sinusoid-like signal with an instantaneous frequency that
is linear with time. A transmitted chirp signal is denoted $s_{tx}(t)$ and is equal to 
\begin{equation}\label{eqn:chirp signal}
s_{tx}(t)=
\begin{cases}
    \exp(j2\pi f_{0}t+j\pi Kt^{2}), & 0\leq t\leq T\\
    0, & \text{otherwise}
\end{cases}
\end{equation}
where $f_{0}$ is the initial frequency, $K$ is the linear FM rate in hertz per second, and $T$ is 
the pulse duration. 

The chirp signal is transmitted several times along the trajectory, and return signals are 
collected for each transmitted signal. Using a ``stop and hop'' approximation, the return radar 
signal is a time shifted, attenuated version of the transmitted signal given by 
\begin{equation}
s_{rx}(t)=As_{tx}(t-\tau)
\end{equation}

The return signal is fed through a matched filter in a process called ``range compression''. The 
matched filter is a time reversed, conjugate version of the transmitted signal $s_{tx}(t)$. The 
output of the matched filter is denoted $s_{out}(t)$ and is equal to the convolution of $s_{rx}(t)$
with $s_{tx}^{*}(T-t)$. 
\begin{align}
    s_{out}(t)& = s_{rx}(t)*s_{tx}^{*}(T-t)\\
    & =
    \begin{cases}
        \int_{0}^{t}s_{rx}(\lambda)s_{tx}^{*}(T-(t-\lambda))d\lambda, & 0\hspace{-1pt}\leq\hspace{-2pt} t\hspace{-2pt}\leq\hspace{-1pt} T\nonumber\\[2pt]
        \int_{t-T}^{T}s_{rx}(\lambda)s_{tx}^{*}(T-(t-\lambda))d\lambda, & T\hspace{-2pt}\leq\hspace{-2pt} t\hspace{-2pt}\leq\hspace{-2pt}2T
    \end{cases}
\end{align}

Evaluating the convolution results in

\begin{equation}
    s_{out}(t)=e^{-j\rho(2\pi f_{0}+\pi KT)}\hspace{-2pt}
    \begin{cases}
        \frac{\sin(\pi K\rho t)}{\pi K\rho}, & 0\hspace{-1pt}\leq\hspace{-2pt} t\hspace{-2pt}<\hspace{-1pt}T\\
        T, & t=T\\
        \frac{\sin(\pi K\rho(2T-t))}{\pi K\rho}, & T\hspace{-2pt}<\hspace{-2pt}t\hspace{-2pt}\leq\hspace{-2pt}2T
    \end{cases}
\end{equation}

where $\rho=T-t$. This expression can be written in a closed form using the sinc function 
$\text{\text{sinc}}(x)=\sin(\pi x)/\pi x$. 

\begin{equation}
    s_{out}(t)=e^{-j\rho(2\pi f_{0}+\pi KT)}\xi\text{sinc}(K\rho\xi)
\end{equation}
where $\xi$ is equal to $T-|t-T|$.

After range compression, the sequential returns from a single target form a hyperbolic curve in the
range compressed data. This hyperbola is quantified via the range equation denoted $R(\boldsymbol{p}_t,\eta)$ 
and defined as
\begin{equation}\label{eqn:time varying range}
    R(\boldsymbol{p}_t,\eta)=\left\Vert \boldsymbol{p}_t-\boldsymbol{p}(\eta)\right\Vert 
\end{equation}
where $\boldsymbol{p}_t$ is the position of a target of the ground and $\boldsymbol{p}$ is the true 
time-varying position of the aircraft from (\ref{eq:cov-veh1}). The aircraft position varies with azimuth 
time (or slow time), $\eta$.

To form an image using BPA, a second matched filter is applied to the range compressed data in the azimuth 
direction. This is called ``azimuth compression''. Azimuth compression
using BPA is performed in the time domain and is dependent on the range equation,
which is dependent on the position of the radar vehicle. For a particular pixel location, $\boldsymbol{p}_{pix}$,
azimuth compression is defined by the summation,
\begin{equation}\label{eqn:azimuth compression}
    A(\boldsymbol{p}_{pix})=\sum_k s_{out}(t_{pix,k})\text{exp}\left\{j4\pi \frac{R_k(\boldsymbol{p}_{pix},\eta)}{\lambda}\right\}
\end{equation}
where $k$ is used to denote the $k^{th}$ range compressed signal, $\lambda$ is the wavelength
at the center frequency of the chirp signal, and $t_{pix,k}$ is the time during the $k^{th}$ 
range compressed signal at which $R_k(\boldsymbol{p}_t,\eta)=R_k(\boldsymbol{p}_{pix},\eta)$.
This time can be calculated via the conversion,
\begin{equation}\label{eqn:time conversion}
    t_{pix} = 2R_k(\boldsymbol{p}_{pix},\eta)/c
\end{equation}
where $c$ is the speed of light. To go from $t_{pix}$ to $t_{pix,k}$, an index in the $k^{th}$
range compressed pulse must be found that corresponds with time $t_{pix}$. Forming a BPA image
is a matter of performing azimuth compression for a collection of pixels within some chosen 
geographical region.

\section{Analysis}\label{sec:analysis}
Errors in the estimated trajectory cause errors in the range equation 
(\ref{eqn:time varying range}), which in turn cause errors in the back-projection equation 
(\ref{eqn:azimuth compression}). The range equation appears in two places in the back-projection 
equation, namely the index of the range compressed data and the phase of the matched filter. As
such, an error in the range equation causes two types of errors. 

First, an error in the index of the range compressed data appears as a change in the hyperbolic 
curve that (\ref{eqn:azimuth compression}) uses to perform azimuth compression. Changes in the 
chosen curve relative to the correct curve manifest as shifts, eccentricity changes, and 
distortions. These errors are referred to as ``curve errors''. Second, an error appears in the
phase of the matched filter. This affects the focus of a target in the final image through a 
phase mismatch. Phase mismatches lead to target blurring. These errors are referred to as 
``phase errors''. 

Curve errors and phase errors are explored individually for position, velocity, and attitude 
navigation errors. Intuition for each type of error is aided by first expanding (\ref{eqn:time varying range})
using a Taylor series approximation. For conciseness, the notation for (\ref{eqn:time varying range})
is abbreviated to $R(\eta)$. According to \cite{cumming_wong}, the
Taylor approximation for the range equation, denoted $\tilde{R}(\eta)$, is approximated as
\begin{equation}\label{eqn:taylor approximation of range}
    \tilde{R}(\eta) \approx R_0 + 
        \frac{d^2\left\Vert R(\eta)\right\Vert^2}{d\eta^2}\bigg\rvert_{\eta=\eta_0}\frac{1}{2R_0}(\eta-\eta_0)^2
\end{equation}
where $\eta_0$ is the time of closest approach and $R_0$ is the range of closest approach, which is
also equal to $R(\eta_0)$. As a common practice in literature, this approximation is expanded about
the time of closest approach $\eta_0$. By doing so, the first order term of the expansion equals zero.

The navigation frame is chosen such that the initial position of the radar platform is the origin.
This origin can be interpreted globally as the point of GPS denial or locally as the beginning of the
synthetic aperture. The platform is assumed to be flying at a constant velocity. For ease of 
visualization in subsequent figures, the radar platform is assumed to be flying northward. In this 
scenario, the true time-varying position of the platform is expressed simply as
\begin{equation}\label{eqn:true position}
    \boldsymbol{p}(\eta)=\boldsymbol{v}_0\eta
\end{equation}
where $\boldsymbol{v}_0$ is the true initial velocity . In (\ref{eqn:time varying range}), the 
time-varying range is expressed in terms of the truth state. Error analysis is performed by replacing
the truth state with the estimated navigation state. Then (\ref{eq:cov-calcEstErrors1}) is used to 
write the navigation state as the difference between the truth state and error state. This is expressed as
\begin{equation}\label{eqn:estimated range equation}
        \hat{R}(\eta)=\left\Vert \boldsymbol{p}_t-(\boldsymbol{p}(\eta)-\delta\boldsymbol{p})\right\Vert 
\end{equation}
where the hat on $\hat{R}(\eta)$ distinguishes this value as an estimate rather than the true value.
This construction allows for an intuitive analysis of the back-projection equation with the help of the
Taylor approximation from (\ref{eqn:taylor approximation of range}), for the cases of position, 
velocity, attitude errors at the beginning of the synthetic aperture.

\subsection{Position Errors}
Using (\ref{eqn:position error equation}), an initial position estimation error, denoted 
$\delta\boldsymbol{p}_0$, is introduced into the estimated range equation.
\begin{equation}\label{eqn:position error range equation}
    \hat{R}(\eta)=\left\Vert \boldsymbol{p}_t-\boldsymbol{v}_0\eta+\delta\boldsymbol{p}_0\right\Vert 
\end{equation}

This equation is then expanded using the Taylor approximation, again denoted with a tilde.
\begin{align}
        \hat{\tilde{R}}(\eta)=&\left\Vert \boldsymbol{p}_t-\boldsymbol{v}_0\eta_0+\delta\boldsymbol{p}_0\right\Vert\\
        &+\frac{ (\boldsymbol{v}_0)^T\boldsymbol{v}_0}{2\left\Vert \boldsymbol{p}_t-\boldsymbol{v}_0\eta_0+\delta\boldsymbol{p}_0\right\Vert}(\eta-\eta_0)^2\nonumber
\end{align}

In the first term of the expansion, $\delta\boldsymbol{p}_0$ causes a constant shift of the hyperbola 
used for azimuth compression.  In the second term, $\delta\boldsymbol{p}_0$ in the denominator is typically
small compared $\boldsymbol{p}_t-\boldsymbol{v}_0\eta_0$. As such, its contribution to the overall error
is very small and can be ignored. In terms of curve errors, the estimated hyperbola is shifted in the
direction of $\delta\boldsymbol{p}_0$ due to the first term of the expansion. For phase errors, constant 
offsets do not affect the overall focus of any imaged target \cite{cumming_wong}. Phase offsets only 
affect knowledge of absolute phase.

The notional effects of position errors are illustrated in Figures 
\ref{fig:cross position illustration}, \ref{fig:along position illustration}, and 
\ref{fig:elevation position illustration}. Each figure is split into three subfigures showing 
how a position error propagates through different stages of radar processing. The first subfigure
shows the error's effect on the flight trajectory. The second subfigure shows the error's effect
on the range compressed data. The third subfigure shows the error's effect on the final image. 

In each figure, light colored or dotted illustrations represent expected data given estimation errors. 
Solid black illustrations represent actual data given no estimation errors. These figures primarily 
provide intuition primarily on curve errors but can be useful in visualizing phase errors as well.

\subsection{Velocity Errors}
\begin{figure*}[p]
    \centering
    \includegraphics[height=0.15\paperheight]{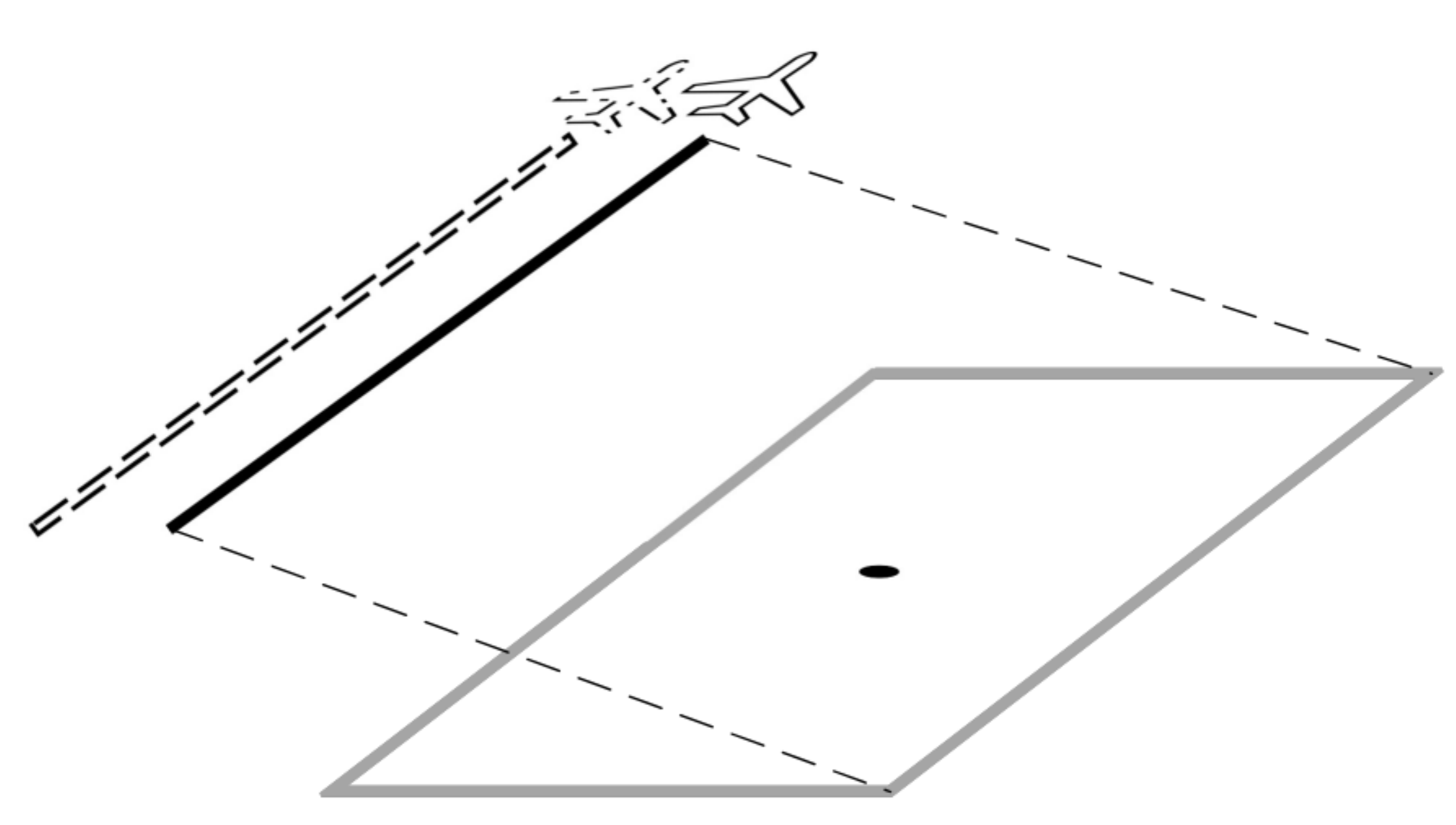}
    \includegraphics[height=0.15\paperheight]{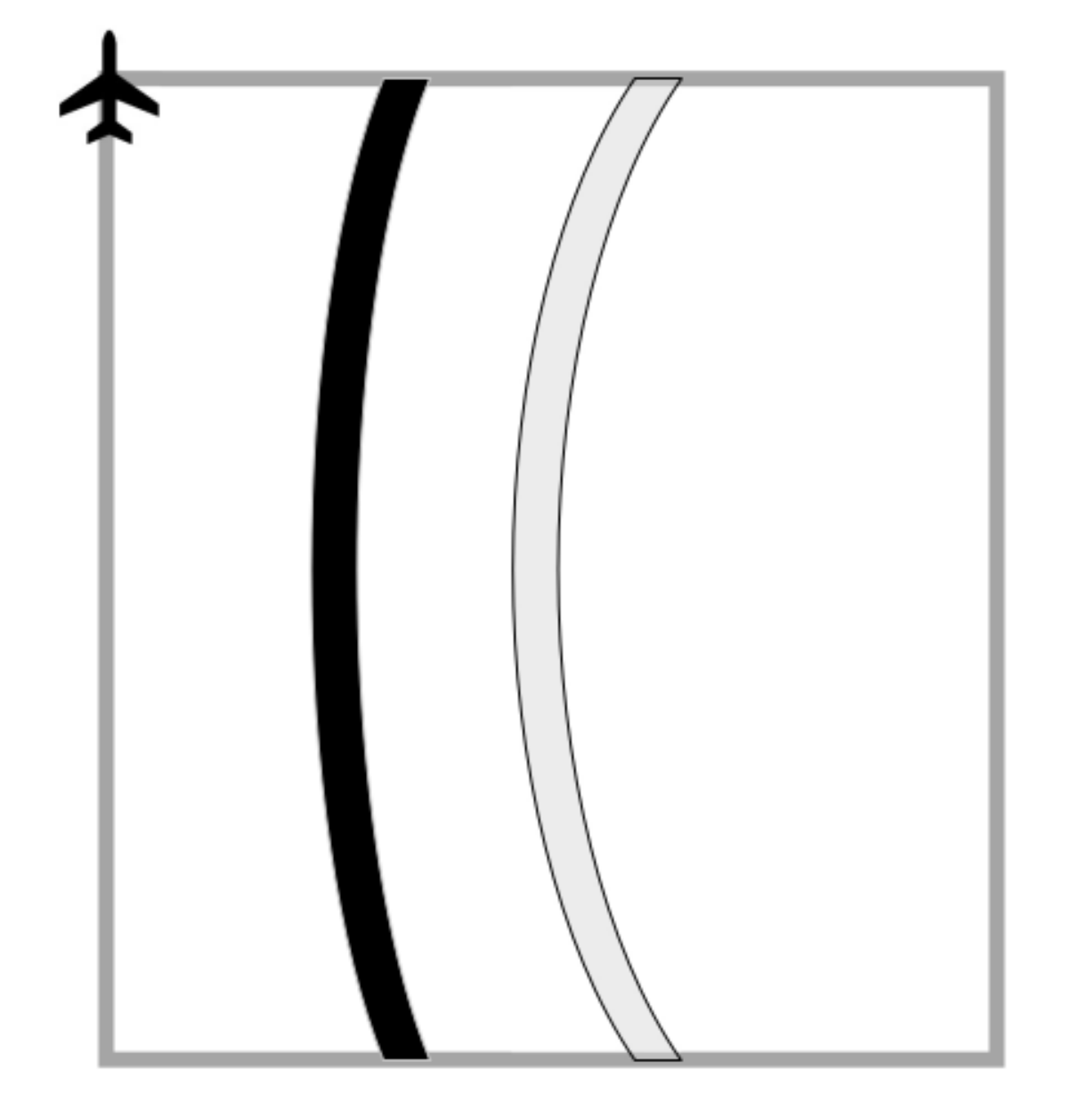}
    \includegraphics[height=0.15\paperheight]{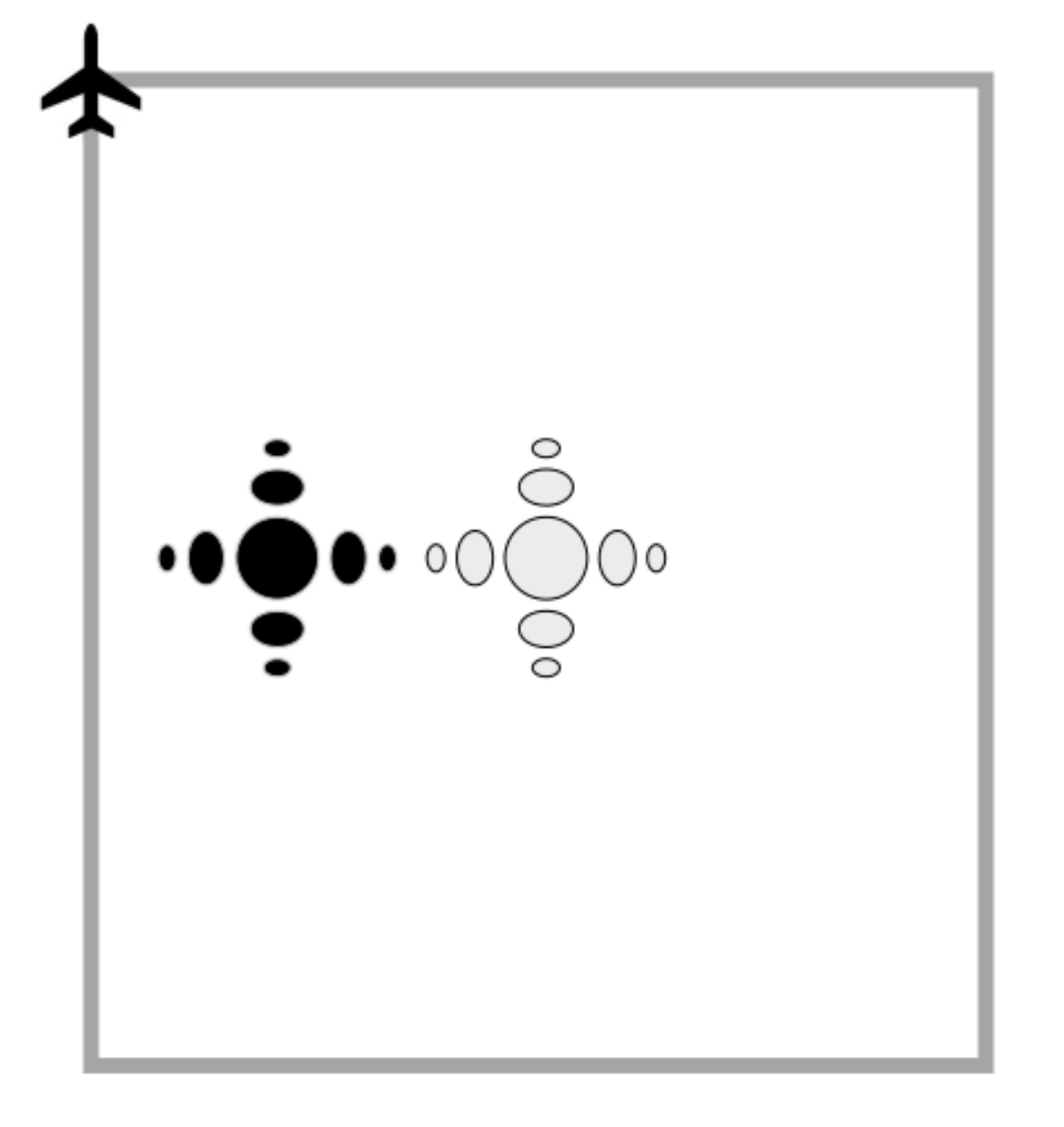}
    \caption{Illustration of how cross track position errors affect the various stages of radar
    processing. Left: flight trajectory with error. Center: range compressed data with error.
    Right: final image with error. Light colored or dotted illustrations represent 
    expected data given estimation errors. Solid black illustrations represent truth data.}
    \label{fig:cross position illustration}
\end{figure*}
\begin{figure*}[p]
    \centering
    \includegraphics[height=0.15\paperheight]{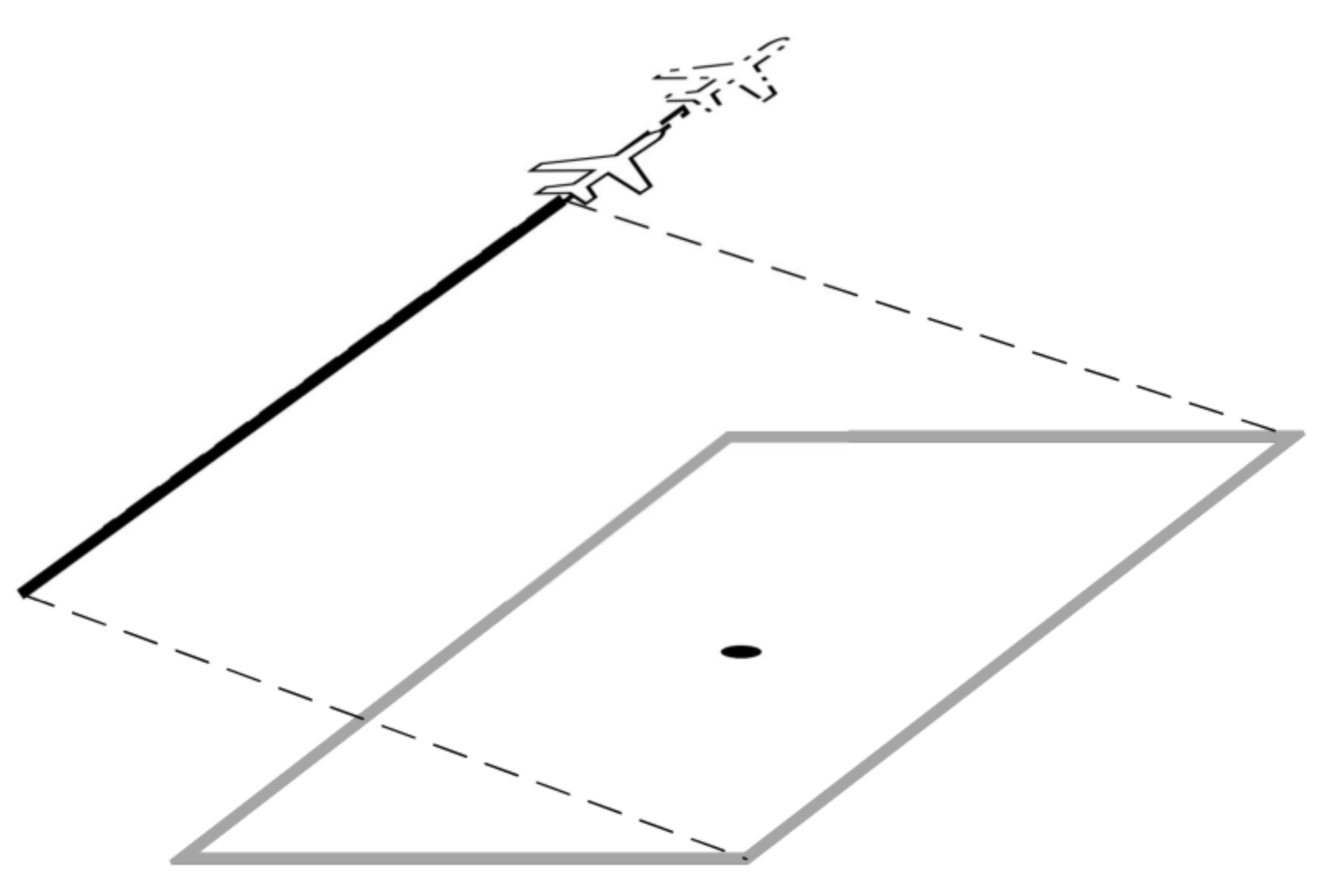}
    \includegraphics[height=0.15\paperheight]{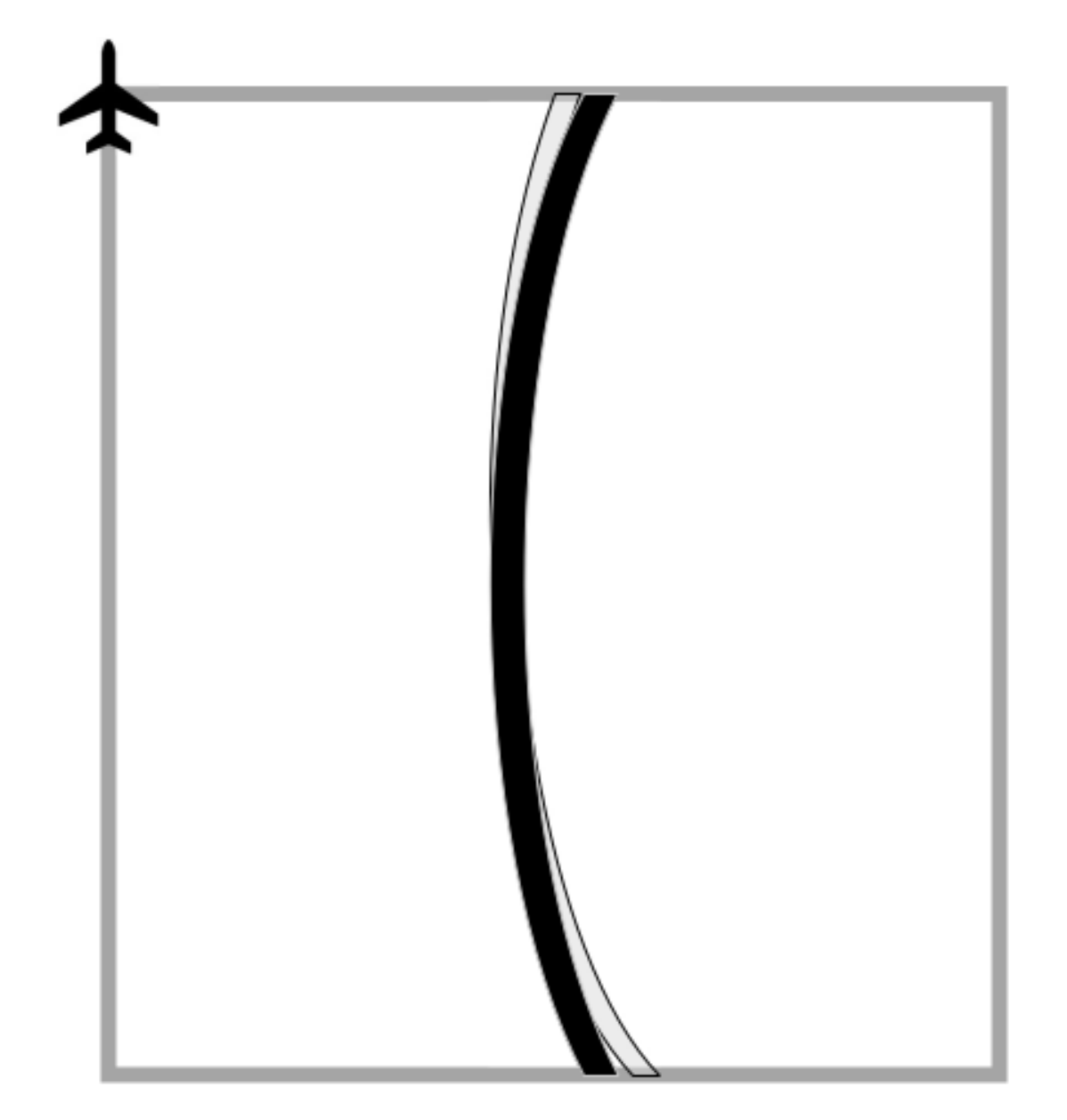}
    \includegraphics[height=0.15\paperheight]{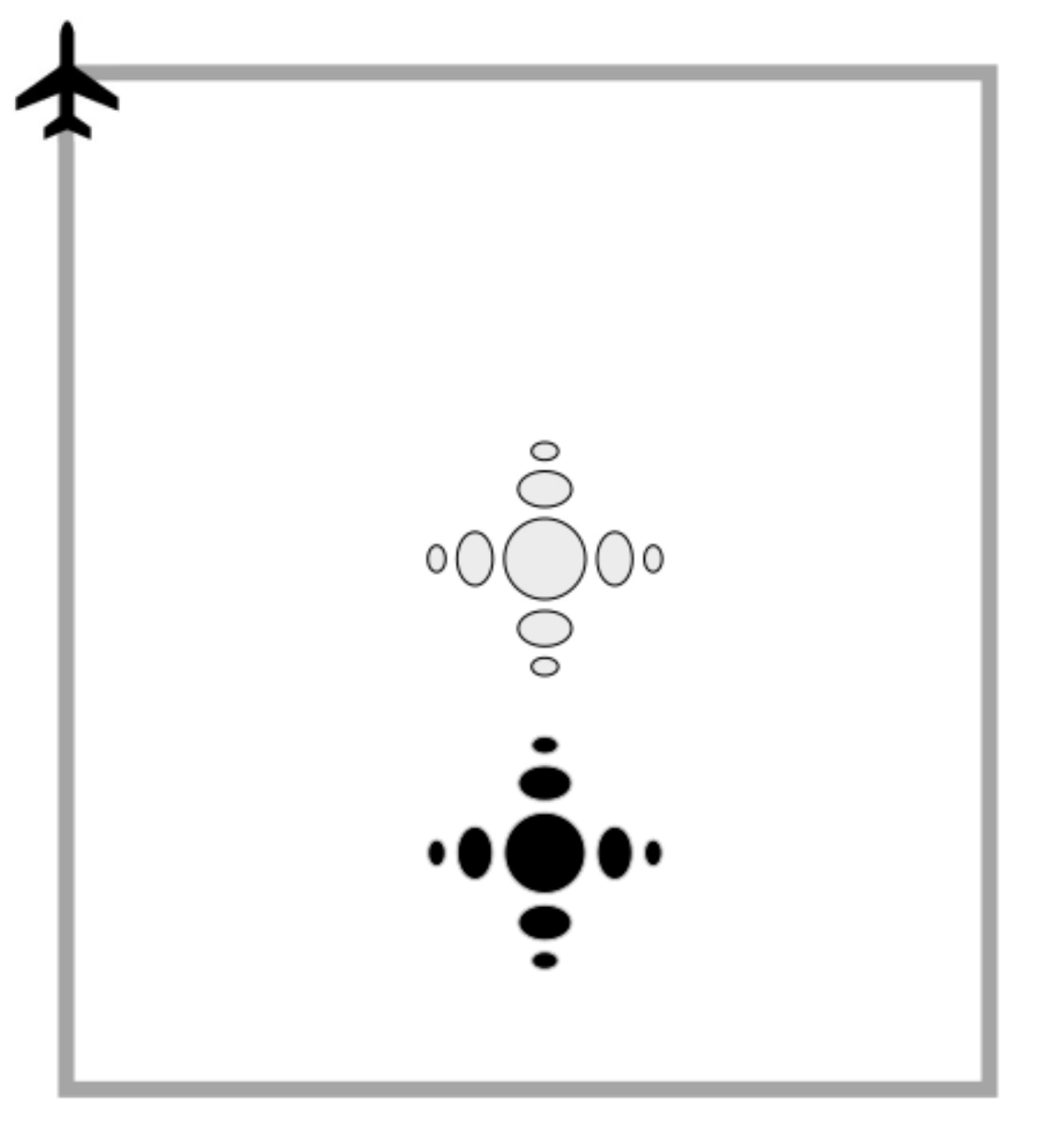}
    \caption{Illustration of how along track position errors affect the various stages of radar
    processing. Left: flight trajectory with error. Center: range compressed data with error.
    Right: final image with error. Light colored or dotted illustrations represent 
    expected data given estimation errors. Solid black illustrations represent truth data.}
    \label{fig:along position illustration}
\end{figure*}

\begin{figure*}[p]
    \centering
    \includegraphics[height=0.15\paperheight]{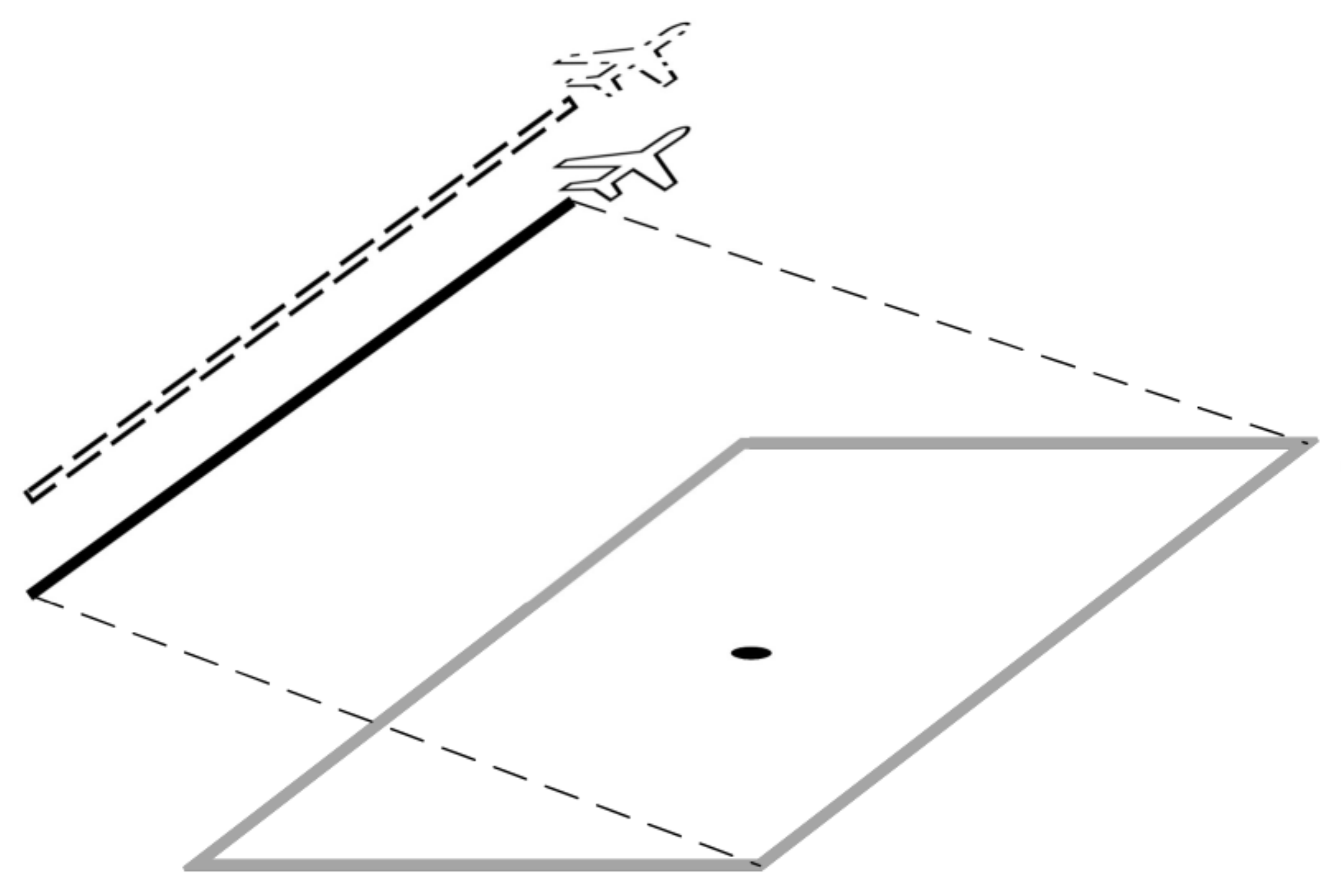}
    \includegraphics[height=0.15\paperheight]{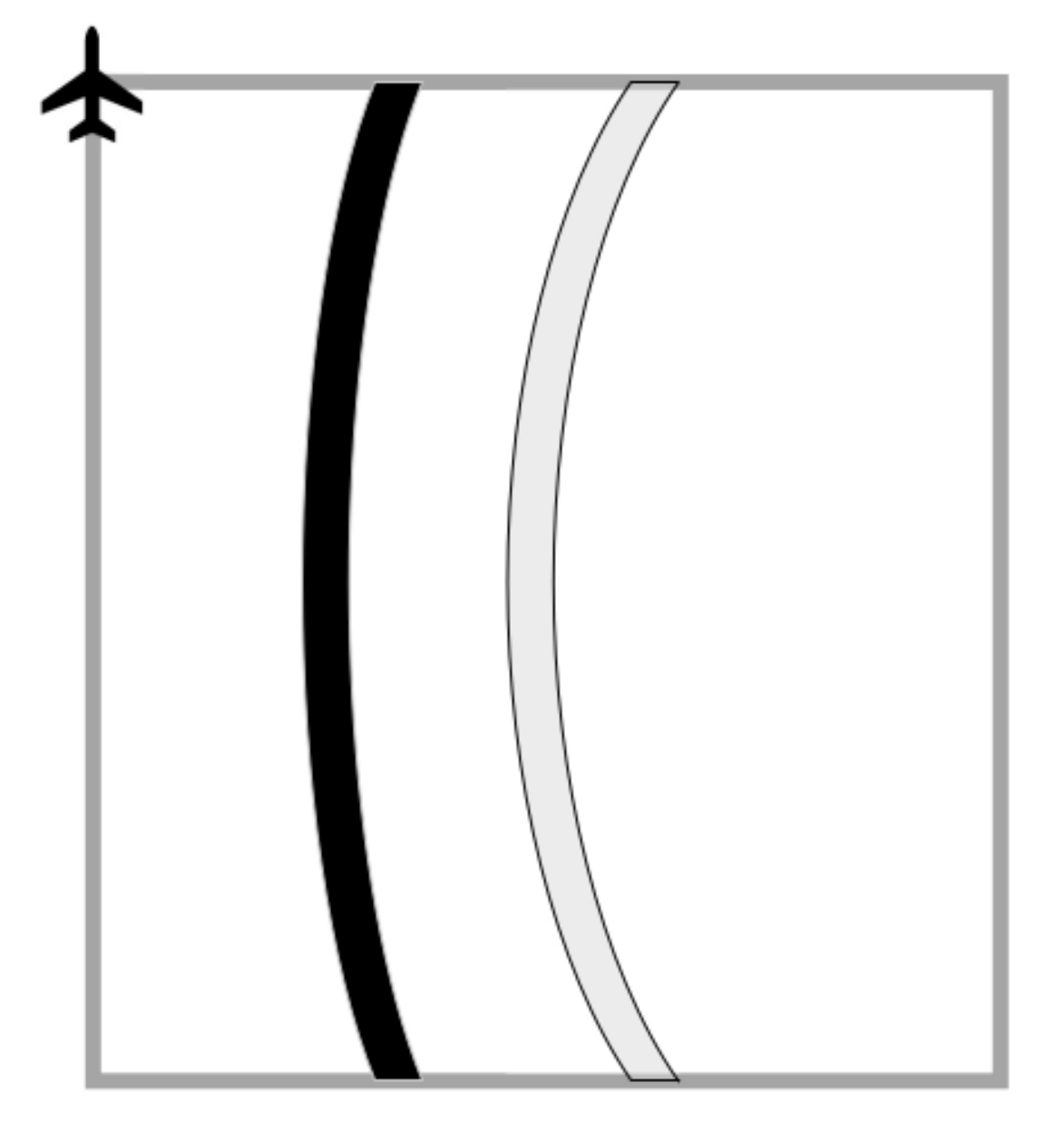}
    \includegraphics[height=0.15\paperheight]{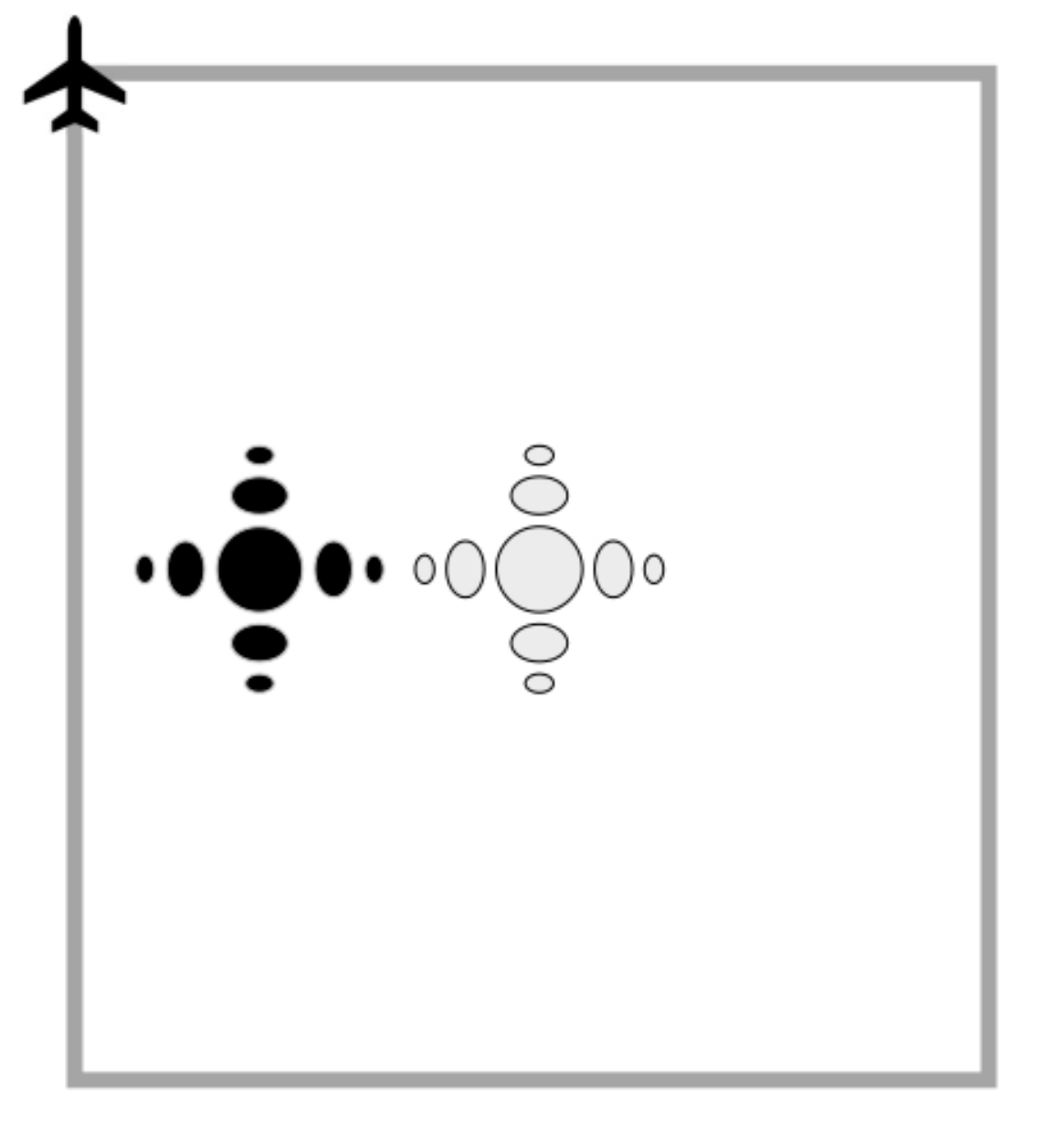}
    \caption{Illustration of how elevation position errors affect the various stages of radar
    processing. Left: flight trajectory with error. Center: range compressed data with error.
    Right: final image with error. Light colored or dotted illustrations represent 
    expected data given estimation errors. Solid black illustrations represent truth data.}
    \label{fig:elevation position illustration}
\end{figure*}

\begin{figure*}[p]
    \centering
    \includegraphics[height=0.15\paperheight]{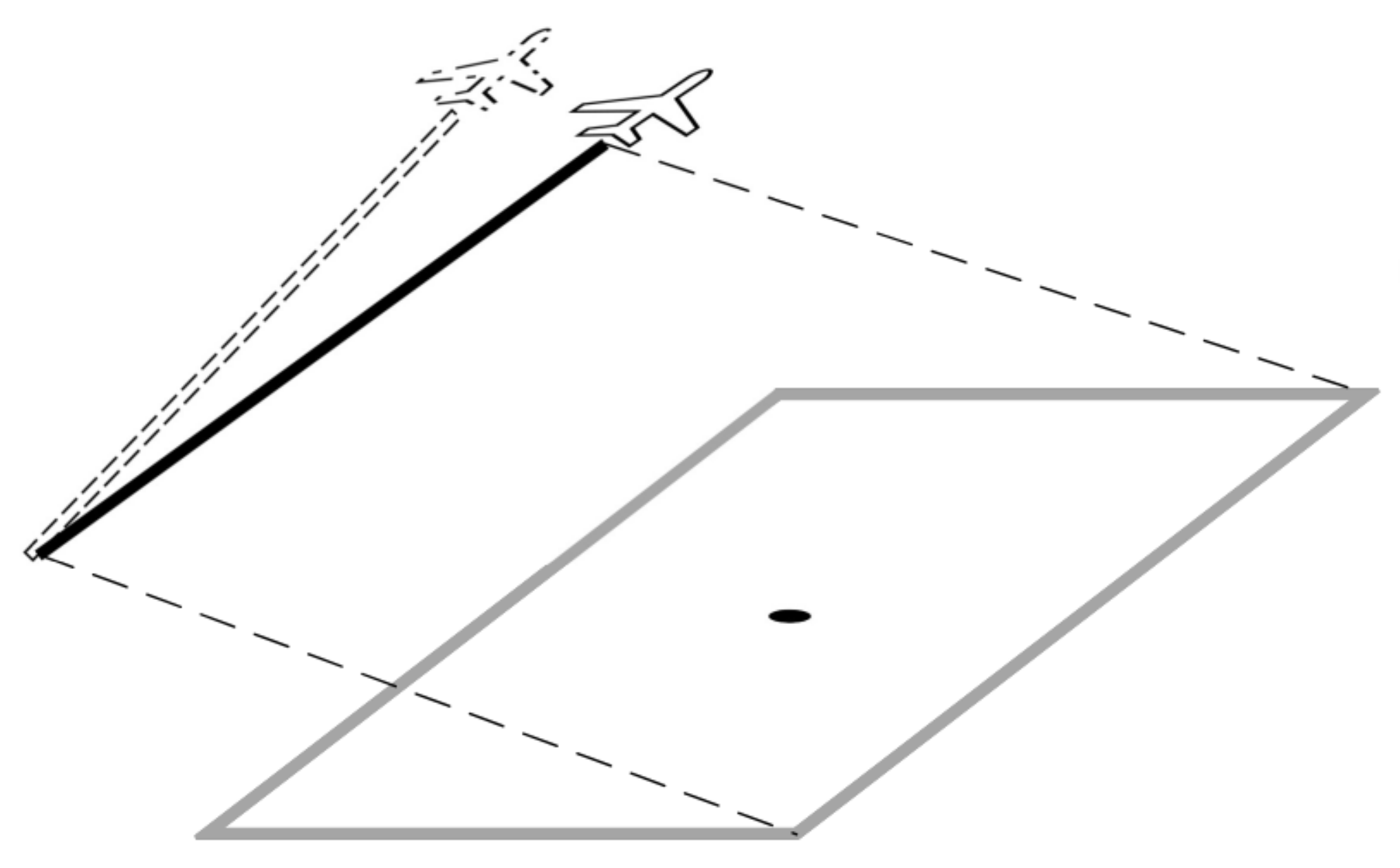}
    \includegraphics[height=0.15\paperheight]{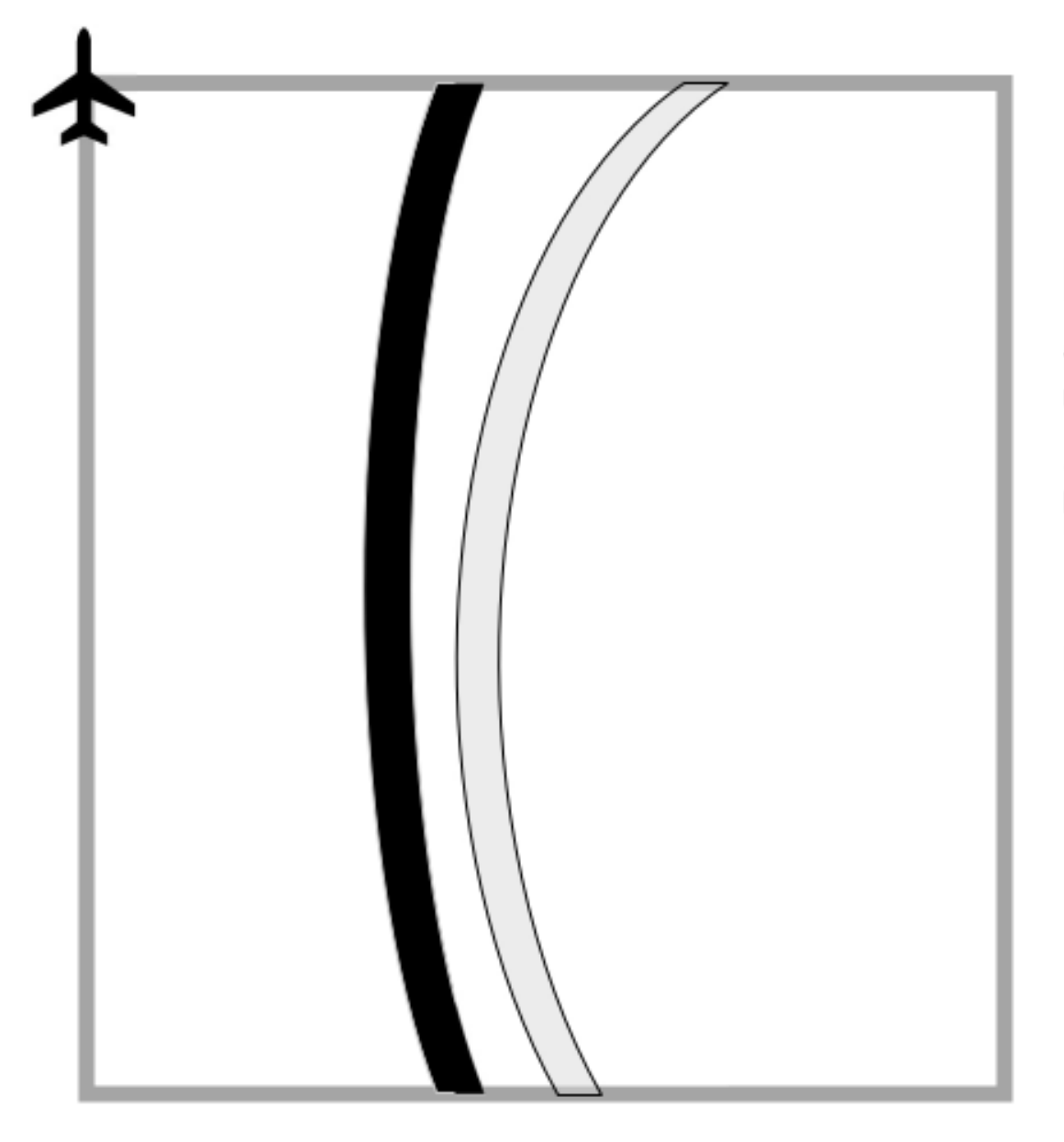}
    \includegraphics[height=0.15\paperheight]{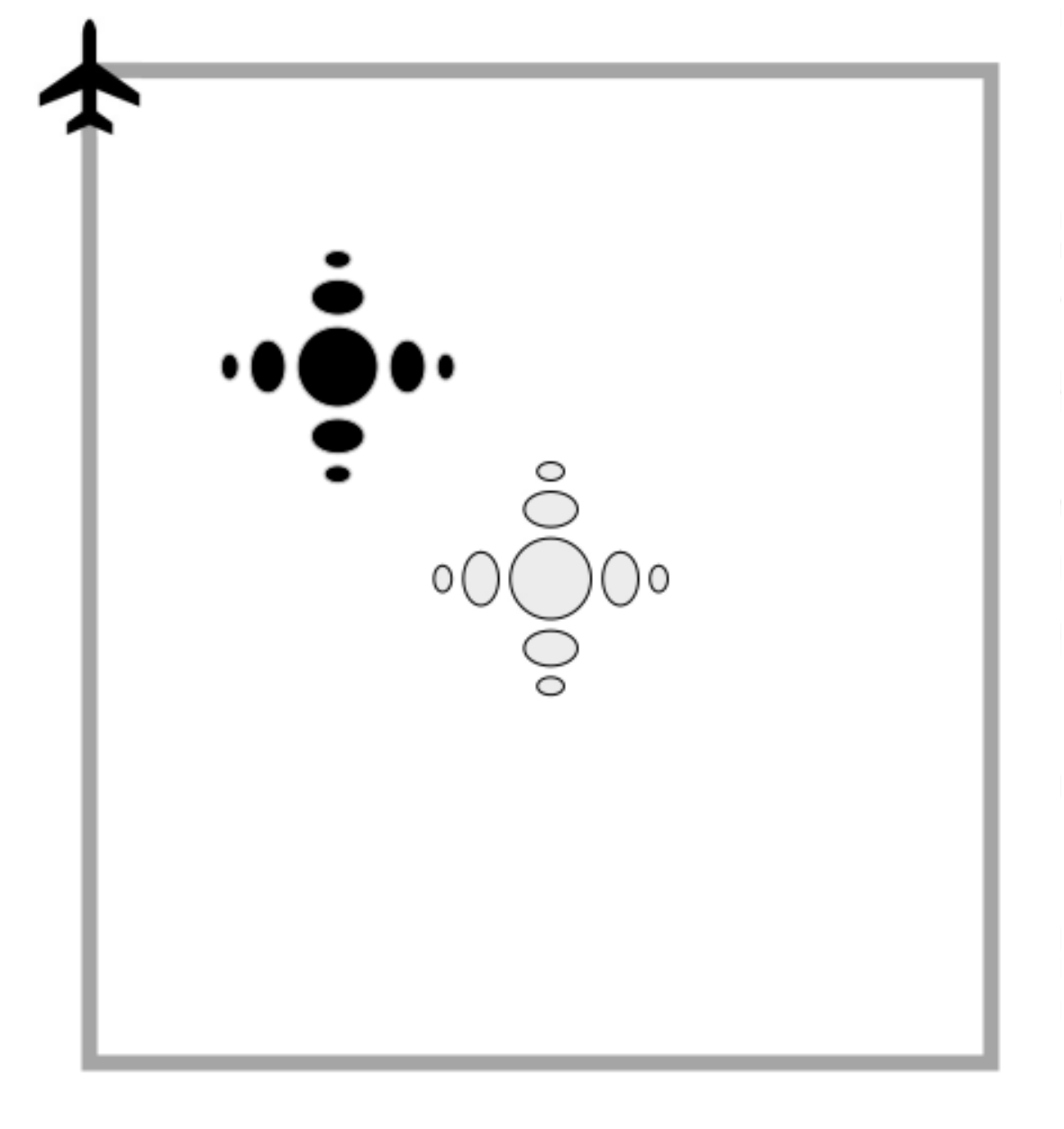}
    \caption{Illustration of how cross track velocity errors affect the various stages of radar
    processing. Left: flight trajectory with error. Center: range compressed data with error.
    Right: final image with error. Light colored or dotted illustrations represent 
    expected data given estimation errors. Solid black illustrations represent truth data.}
    \label{fig:cross_vel}
\end{figure*}

\begin{figure*}[p]
    \centering
    \includegraphics[height=0.15\paperheight]{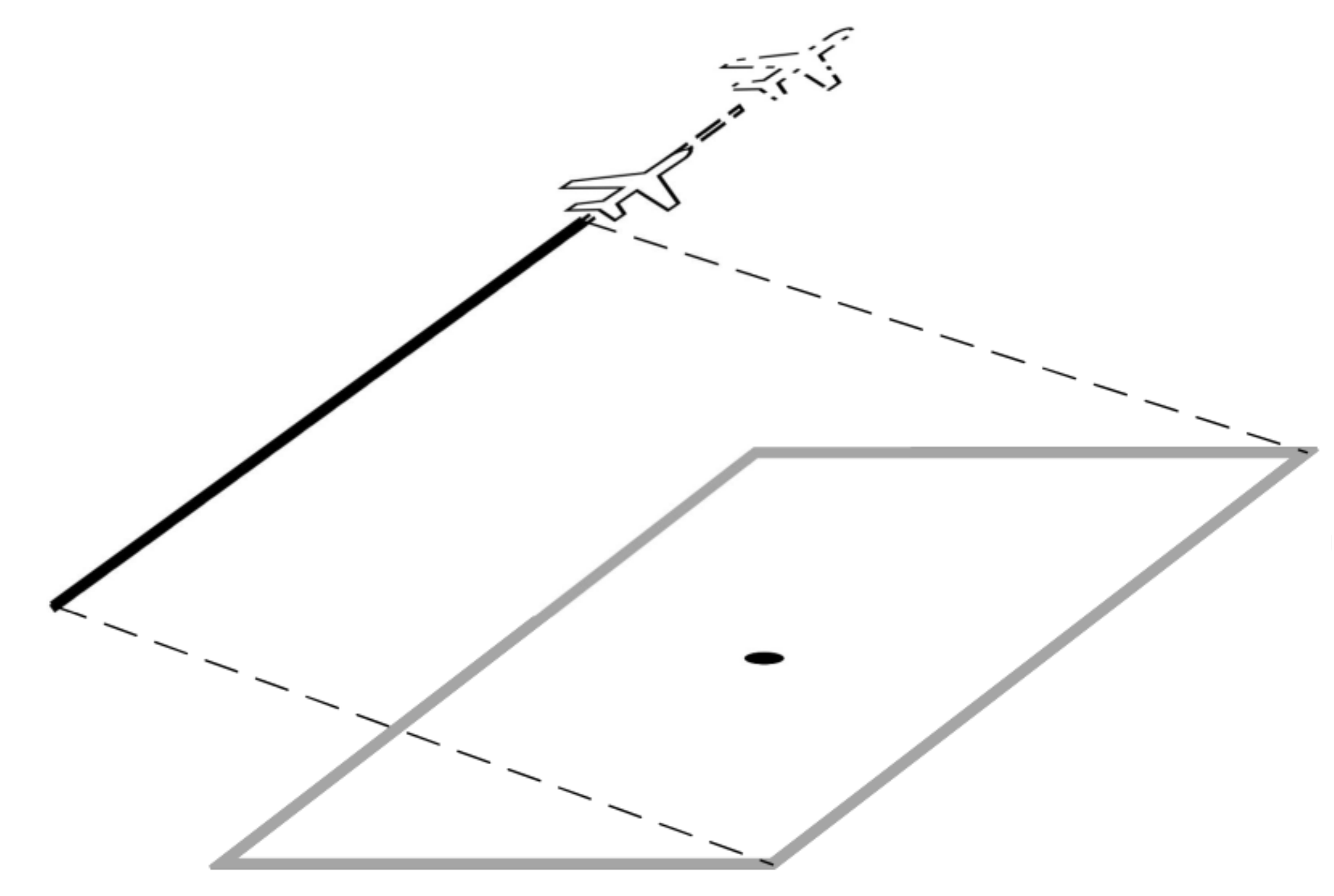}
    \includegraphics[height=0.15\paperheight]{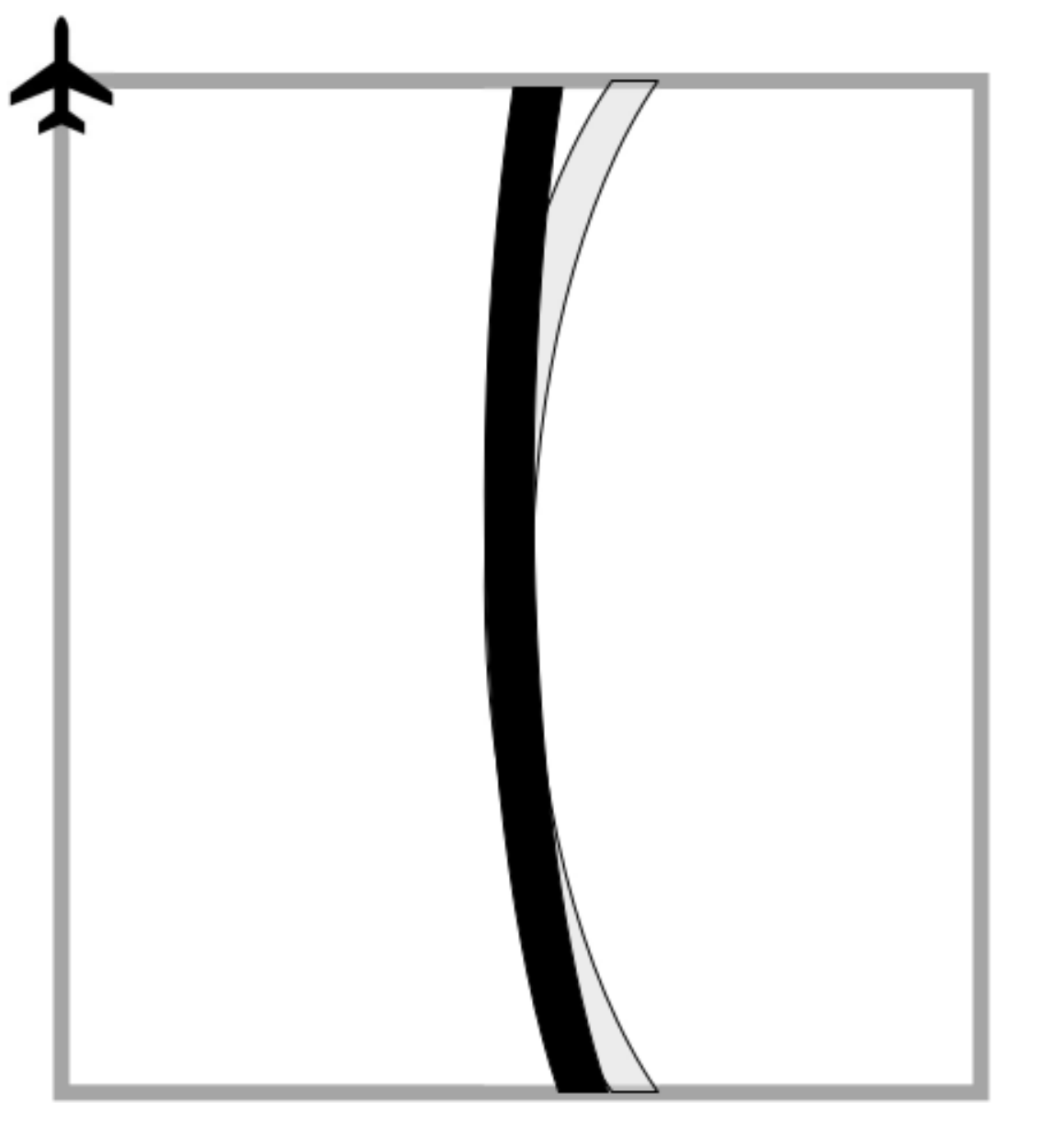}
    \includegraphics[height=0.15\paperheight]{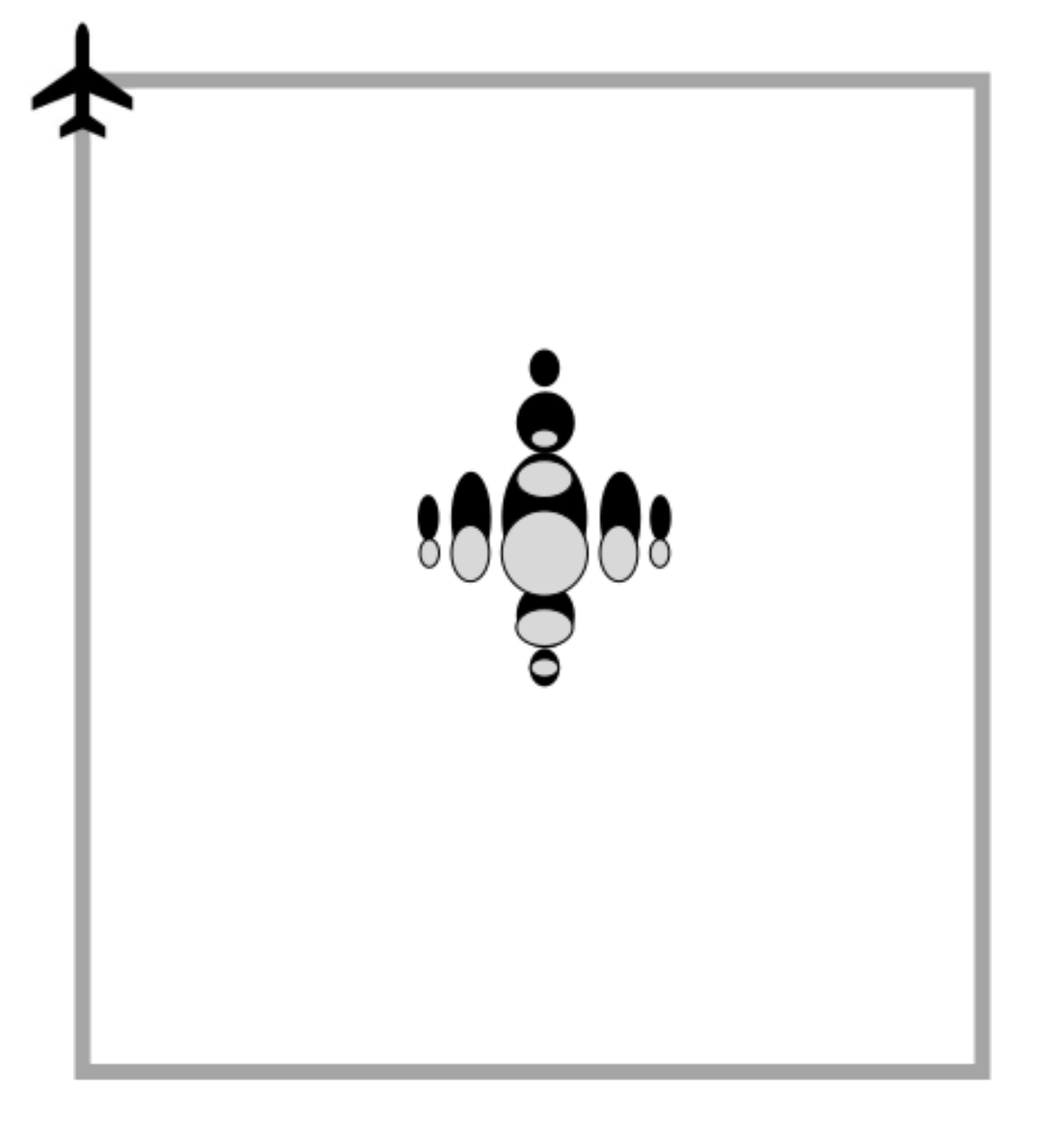}
    \caption{Illustration of how along track velocity errors affect the various stages of radar
    processing. Left: flight trajectory with error. Center: range compressed data with error.
    Right: final image with error. Light colored or dotted illustrations represent 
    expected data given estimation errors. Solid black illustrations represent truth data.}
    \label{fig:along_vel}
\end{figure*}

\begin{figure*}[p]
    \centering
    \includegraphics[height=0.15\paperheight]{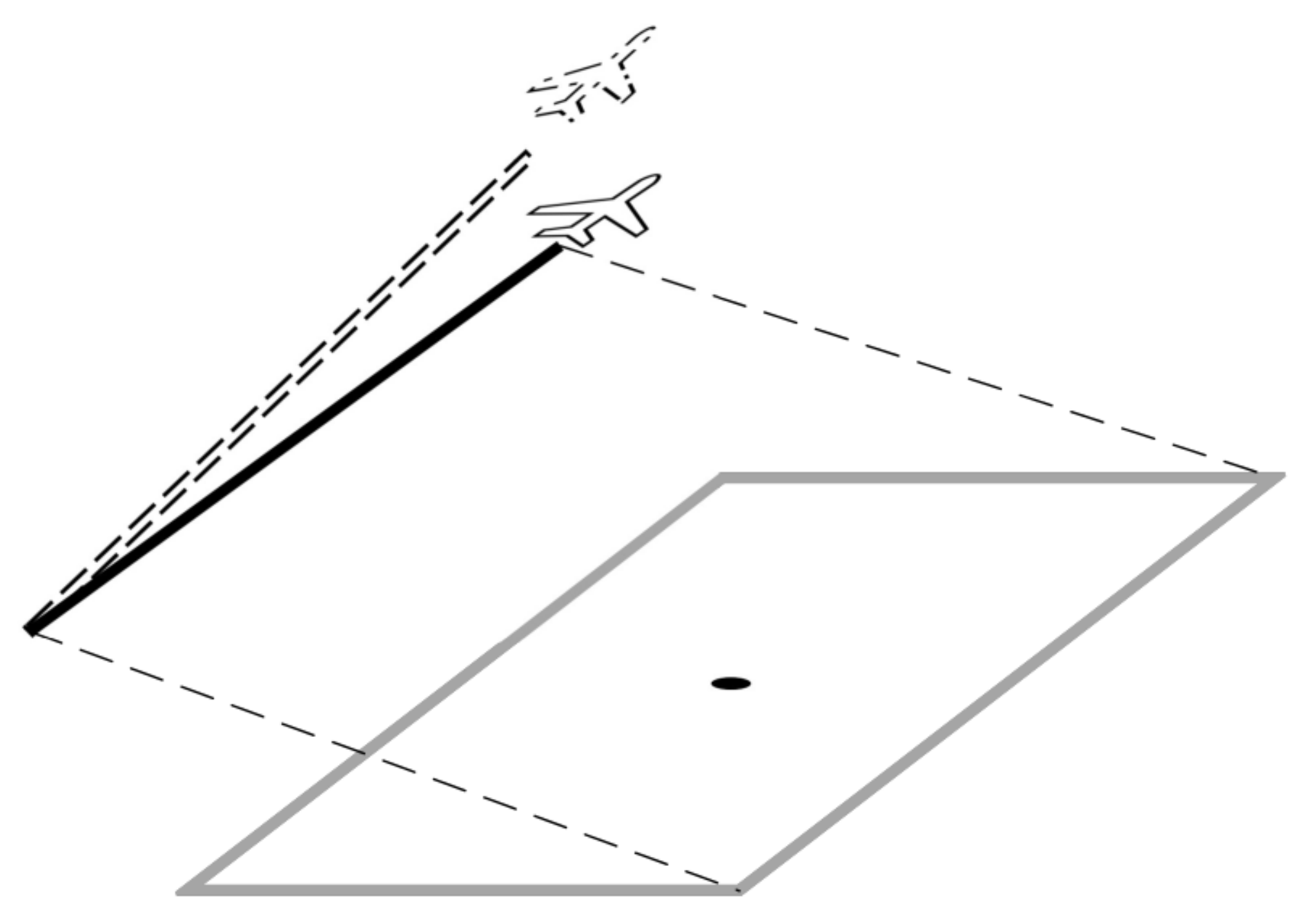}
    \includegraphics[height=0.15\paperheight]{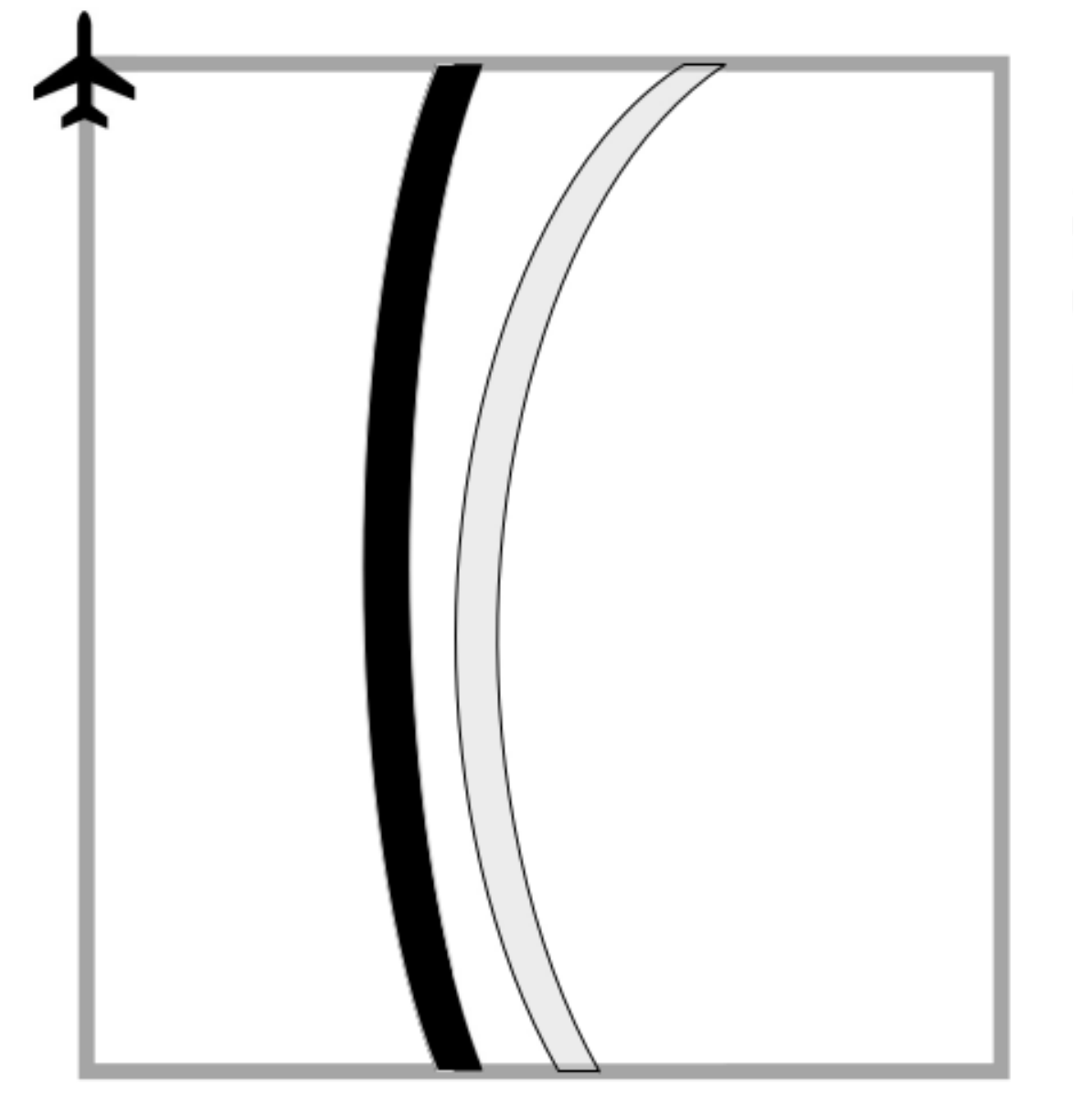}
    \includegraphics[height=0.15\paperheight]{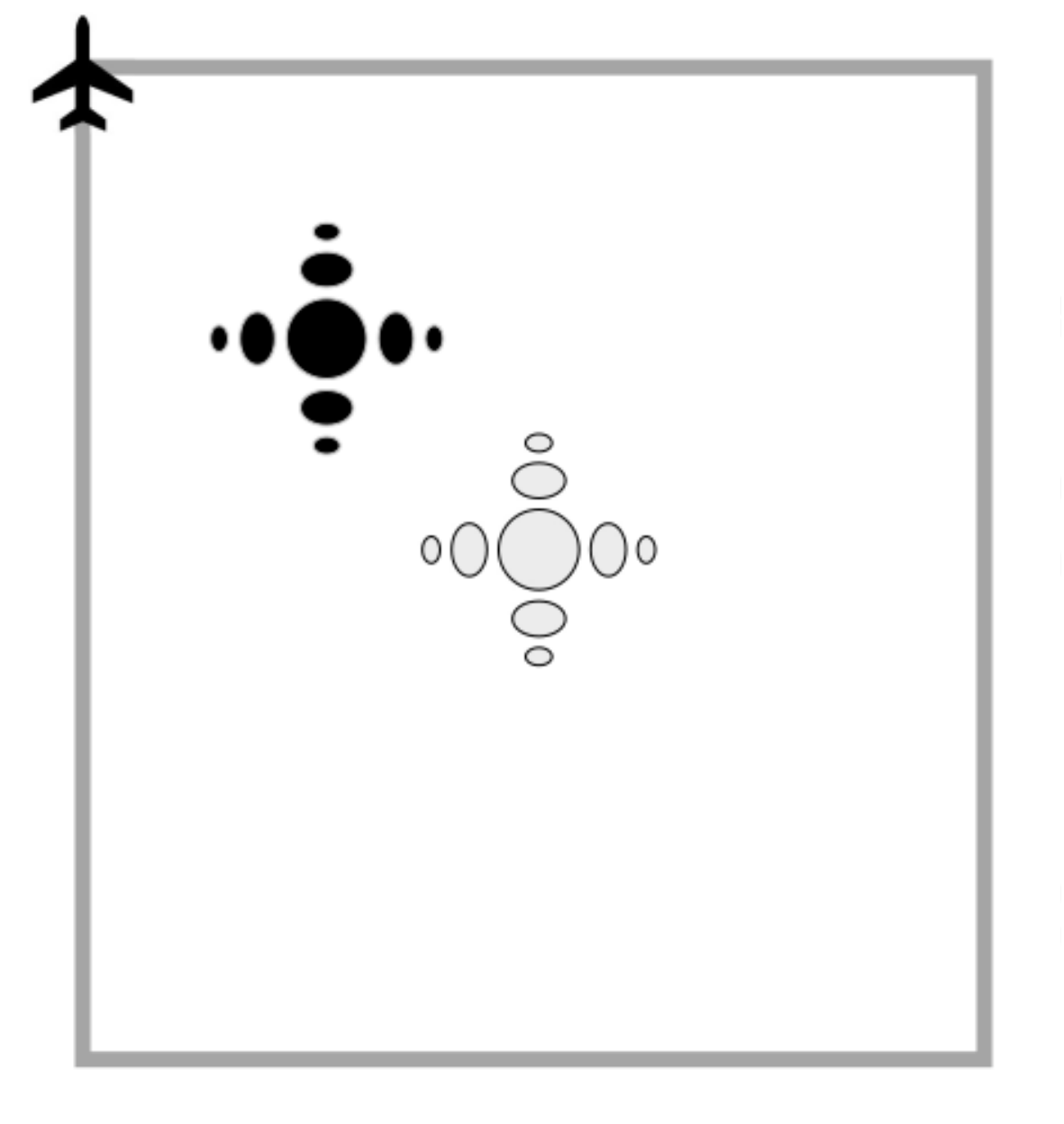}
    \caption{Illustration of how elevation velocity errors affect the various stages of radar
    processing. Left: flight trajectory with error. Center: range compressed data with error.
    Right: final image with error. Light colored or dotted illustrations represent 
    expected data given estimation errors. Solid black illustrations represent truth data.}
    \label{fig:elev_vel}
\end{figure*}

\begin{figure*}[p]
    \centering
    \includegraphics[height=0.15\paperheight]{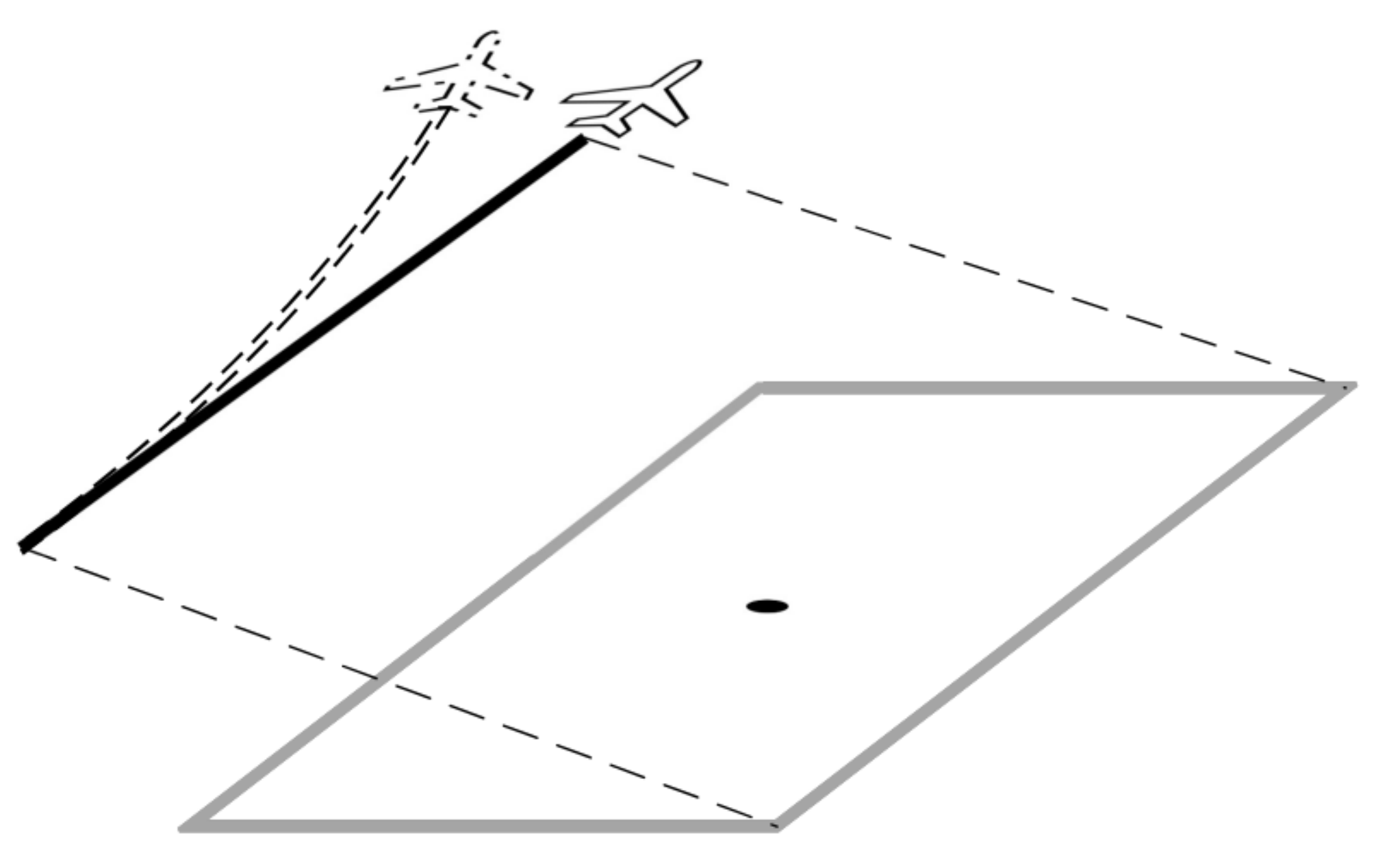}
    \includegraphics[height=0.15\paperheight]{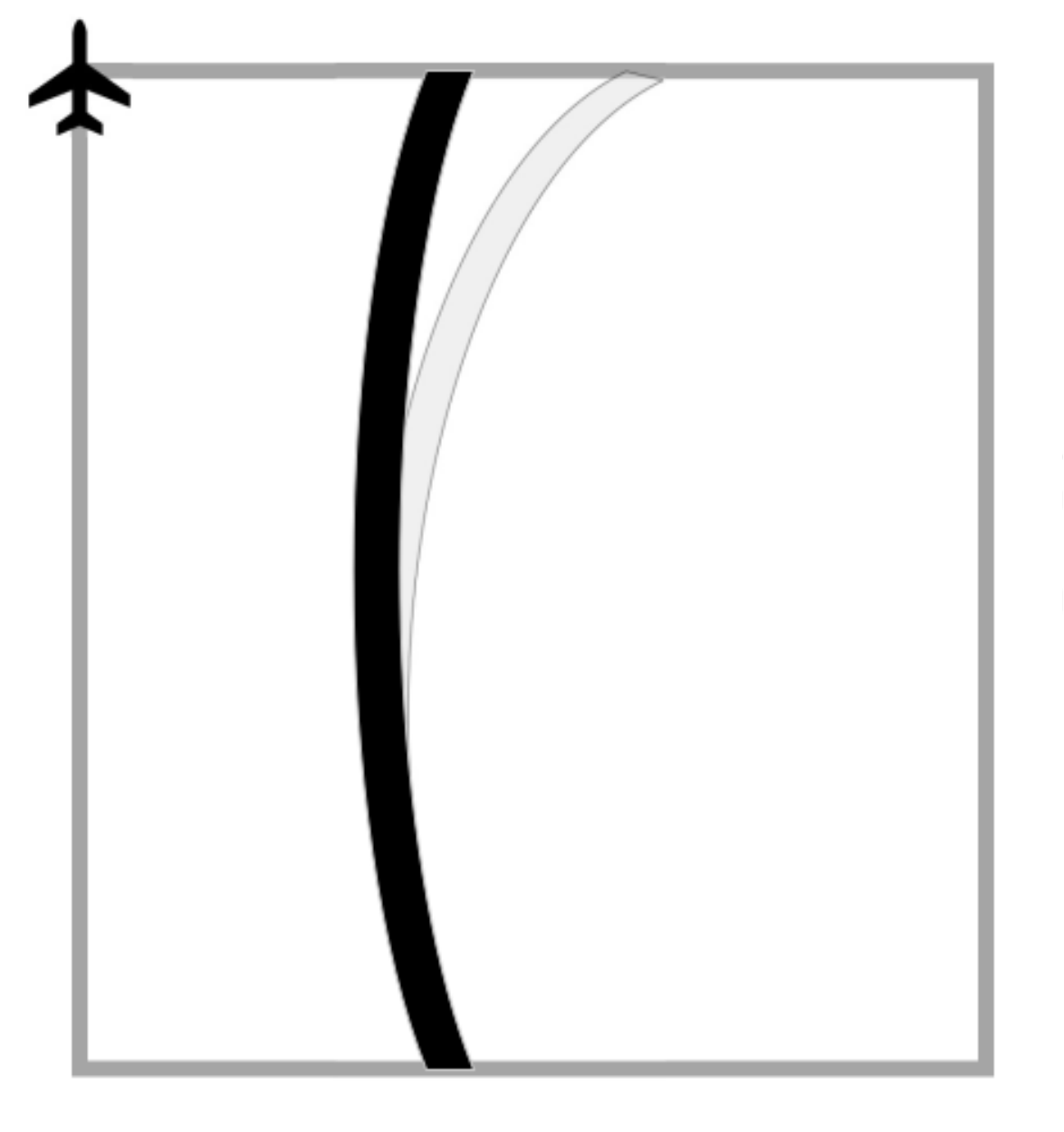}
    \includegraphics[height=0.15\paperheight]{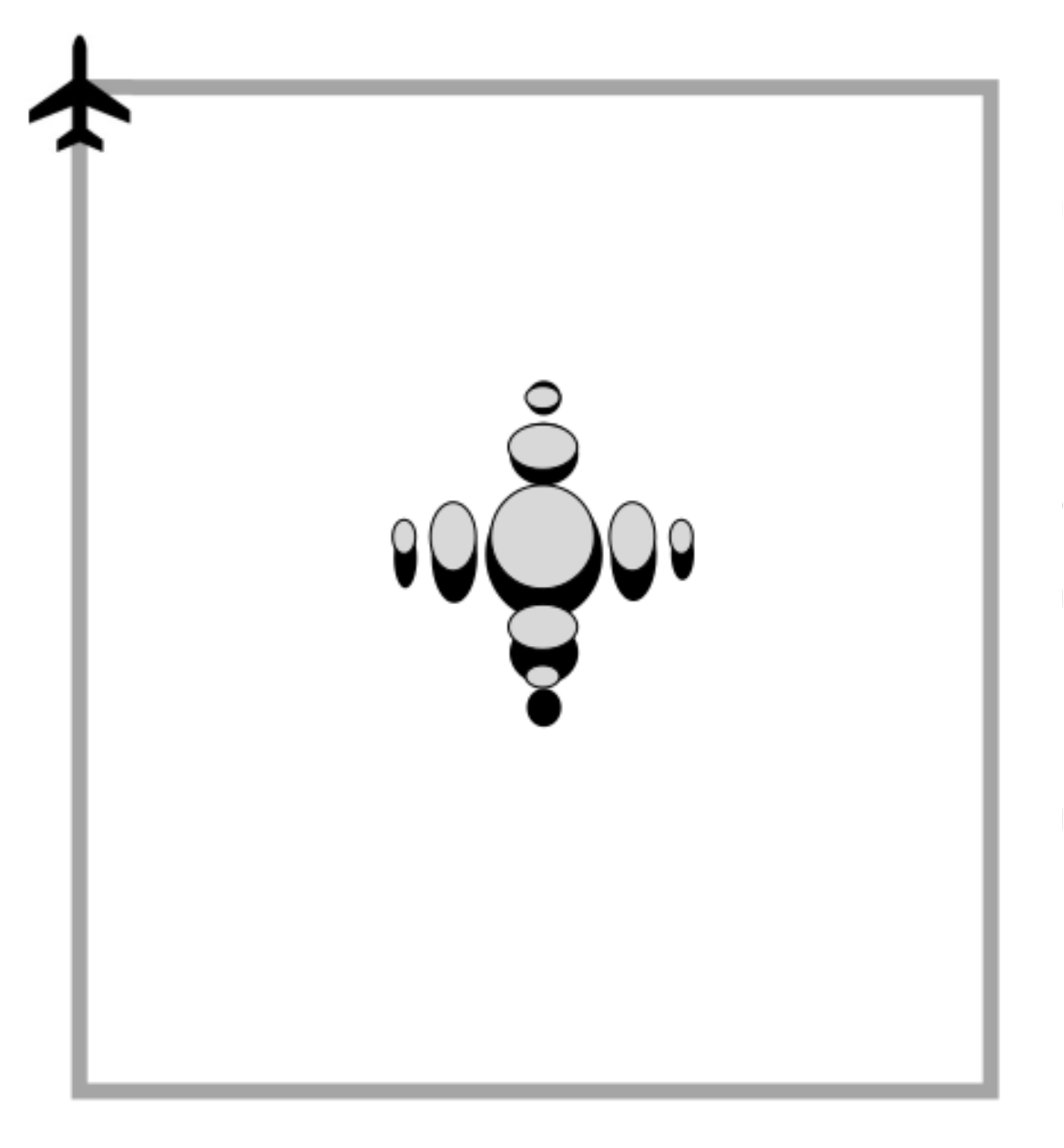}
    \caption{Progression of roll errors through the SAR data. Left: flight trajectory with error.
    Center: range compressed data with error. Right: final image with error. Light colored or 
    dotted illustrations represent expected data given estimation errors. Solid black 
    illustrations represent truth data.}
    \label{fig:roll_error}
\end{figure*}

\begin{figure*}[p]
    \centering
    \includegraphics[height=0.15\paperheight]{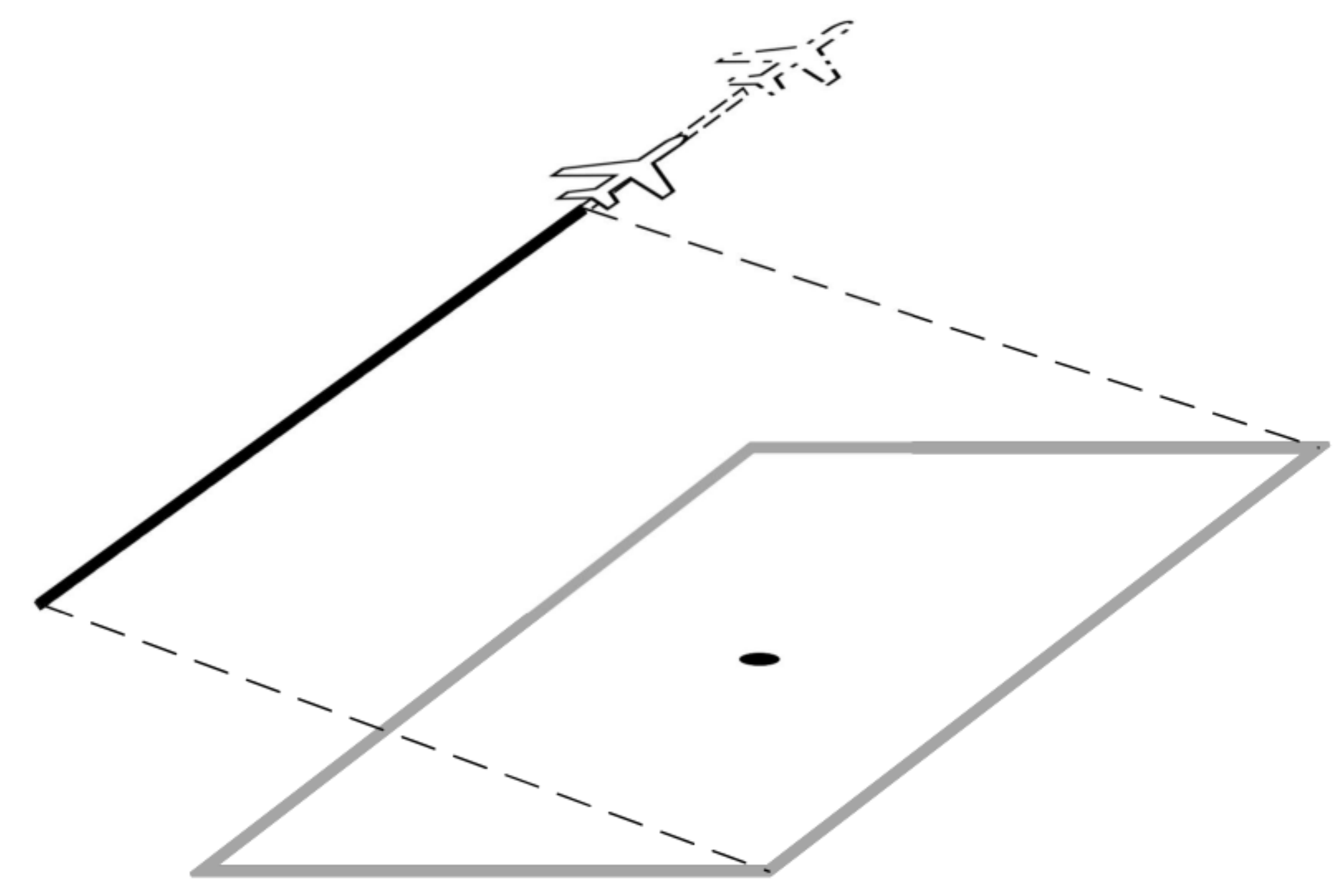}
    \includegraphics[height=0.15\paperheight]{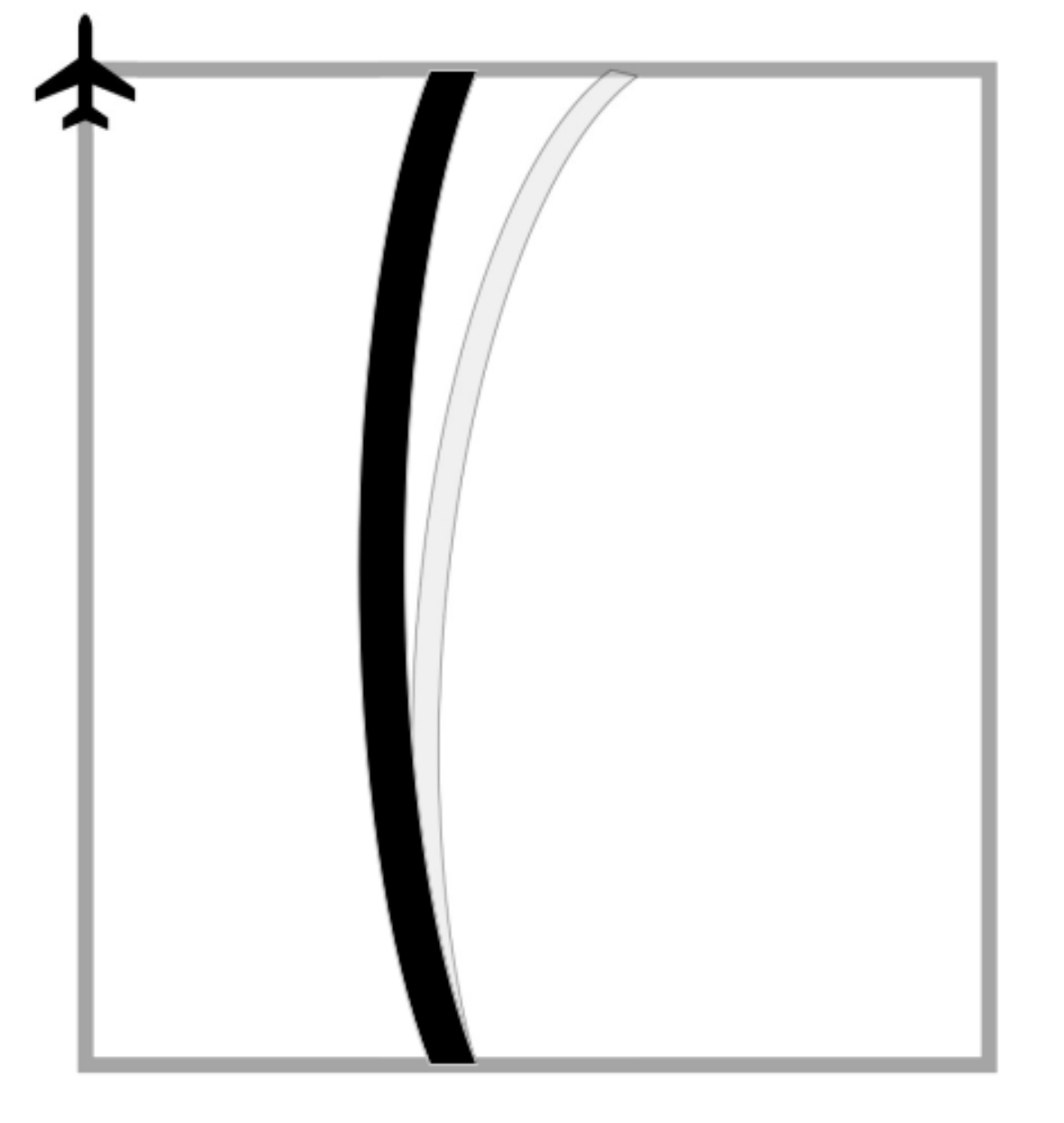}
    \includegraphics[height=0.15\paperheight]{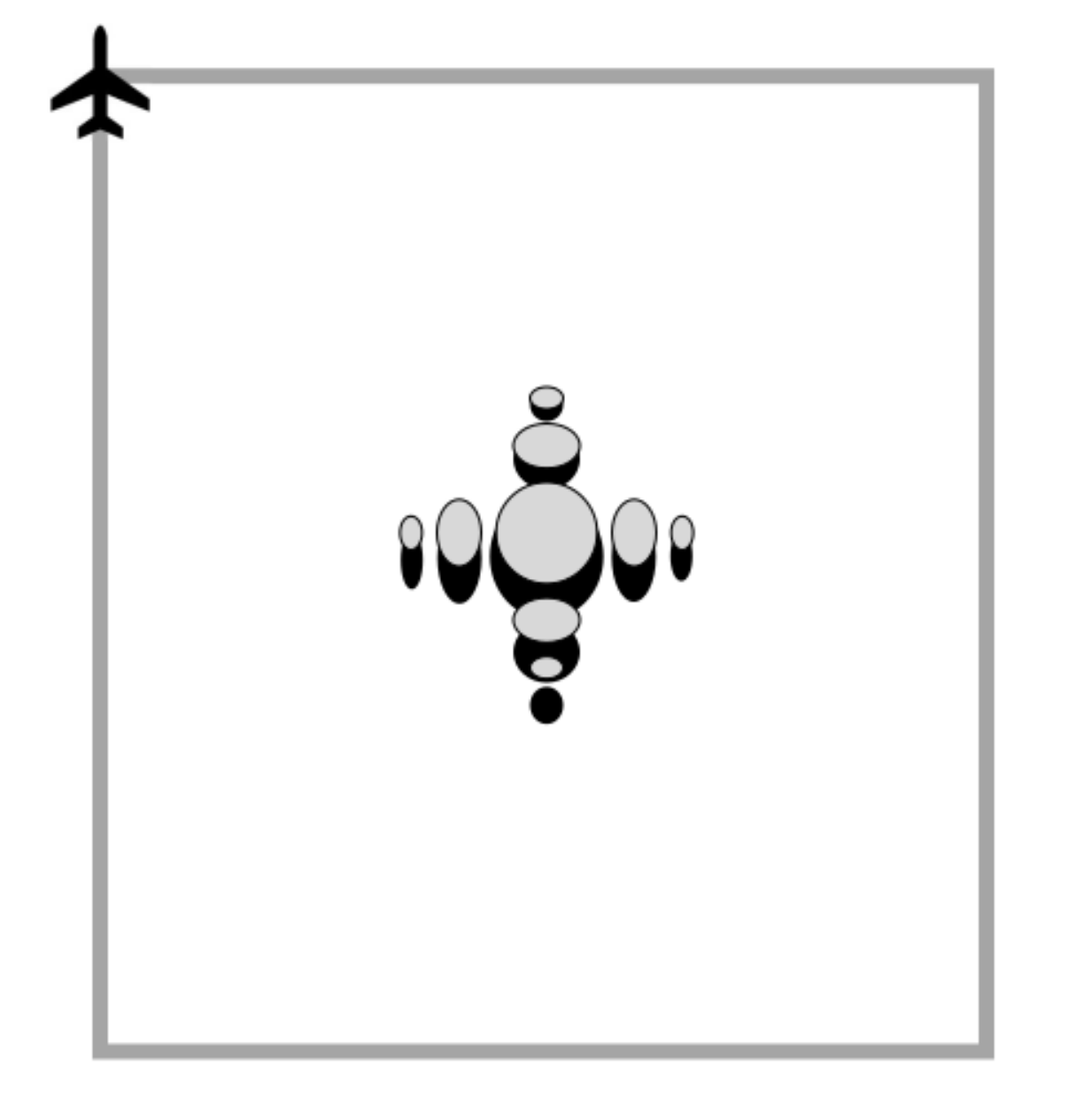}
    \caption{Progression of pitch error through the SAR data. Left: flight trajectory with error.
    Center: range compressed data with error. Right: final image with error. Light colored or 
    dotted illustrations represent expected data given estimation errors. Solid black 
    illustrations represent truth data.}
    \label{fig:pitch_error}
\end{figure*}
From (\ref{eqn:position error equation}), an initial velocity estimation error is introduced into the 
estimated range equation as
\begin{equation}\label{velocity error range equation}
    \hat{R}(\eta)=\left\Vert \boldsymbol{p}_t-(\boldsymbol{v}_0-\delta\boldsymbol{v}_0)\eta\right\Vert 
\end{equation}

Again, the Taylor expansion is taken and results in
\begin{align}
    \hat{\tilde{R}}(\eta) =& \left\Vert \boldsymbol{p}_t-(\boldsymbol{v}_0-\delta\boldsymbol{v}_0)\eta_0\right\Vert\\
    &+\frac{(\boldsymbol{v}_0)^T\boldsymbol{v}_0-2(\boldsymbol{v}_0)^T\delta\boldsymbol{v}_0+(\delta\boldsymbol{v}_0)^T\delta\boldsymbol{v}_0}
    {\left\Vert \boldsymbol{p}_t-(\boldsymbol{v}_0-\delta\boldsymbol{v}_0)\eta_0\right\Vert }
    (\eta-\eta_0)^2\nonumber
\end{align}

In the first term of the expansion, there is again a constant offset due to $\delta\boldsymbol{v}_0$.
As is the case for initial position errors, this constant offset causes a shifted curve error
and a negligible offset phase error. In the second term of the expansion, 
$(\delta\boldsymbol{v}_0)^T\delta\boldsymbol{v}_0$ in the numerator is a quadratic error term and 
contribute little overall error. Similar to the case of position errors, the $\delta\boldsymbol{v}_0\eta$
term in the denominator contributes negligible overall error due to its relative size compared to the
rest of the denominator. 

In the numerator of the second term, $2(\boldsymbol{v}_0)^T\delta\boldsymbol{v}_0$
causes a time-varying error. In terms of curve errors, this changes the eccentricity of the expected
hyperbola in the range compressed data. For phase errors, this term can be thought of a linearly changing 
frequency error or azimuth FM rate error. 

An azimuth FM rate error is characterized by a phase that changes
quadratically in time. A quadratically varying phase yields a linearly changing instantaneous frequency.
This is similar to the linear FM signal modeled by equation (\ref{eqn:chirp signal}). For both curve errors
and phase errors, the second numerator term results in blurring of the imaged target in the azimuth
dimension. This blur is only present in along track errors, as cross track and elevation errors result in
a $\delta\boldsymbol{v}_0$ that is orthogonal to $\boldsymbol{v}_0$.

Again, the notional effects are illustrated in Figures \ref{fig:cross_vel}, \ref{fig:along_vel}, and
\ref{fig:elev_vel} for various stages of SAR processing. Each figure is again split into three subfigures 
with identical interpretations as Figures \ref{fig:cross position illustration}, 
\ref{fig:along position illustration}, and \ref{fig:elevation position illustration}.

\subsection{Attitude Errors}
Again using (\ref{eqn:position error equation}), initial attitude errors are injected into the range 
equation to yield
\begin{equation}\label{attitude error range equation}
    \hat{R}(\eta)=\left\Vert \boldsymbol{p}_t-\boldsymbol{v}_0\eta+
        \boldsymbol{\nu}^{n}\times\delta\boldsymbol{\theta}_0\frac{\eta^2}{2}\right\Vert 
\end{equation}
Errors in attitude manifest as errors in acceleration. Specific effects from attitude errors are 
apparent after computing the cross product. Using constant accelerometer measurements,
\begin{equation}\label{eqn:acceleration errors}
    \boldsymbol{\nu}^{n}\times\delta\boldsymbol{\theta}_0 = 
    \left[
    \begin{array}{ccc}
        0\\
        0\\
        -g
    \end{array}
    \right]\times
    \left[
    \begin{array}{ccc}
        \delta\theta_{x,0}\\
        \delta\theta_{y,0}\\
        \delta\theta_{z,0}
    \end{array}
    \right] = 
    \left[
    \begin{array}{ccc}
        \delta\theta_{y,0}g\\
        -\delta\theta_{x,0}g\\
        0
    \end{array}
    \right]
\end{equation}
This equation illustrates how attitude errors only cause acceleration errors in the along track and cross
track directions. Specifically, errors in roll, $\delta\theta_{x,0}$, cause cross track
acceleration errors. Errors in pitch, $\delta\theta_{y,0}$, cause along track acceleration
errors. Errors in yaw do not cause any errors in acceleration. 

For conciseness, acceleration errors resulting from (\ref{eqn:acceleration errors}) are
collectively referred to as $\delta\dot{\boldsymbol{v}}_0$. As such, the estimated range equation
takes the form
\begin{equation}\label{acceleration error range equation}
    \hat{R}(\eta)=\left\Vert \boldsymbol{p}_t-\boldsymbol{v}_0\eta+
    \frac{1}{2}\delta\boldsymbol{\dot{v}}_0\eta^2\right\Vert 
\end{equation}

The effects of acceleration errors are again explored using the Taylor approximation of the estimated
range equation.
\begin{align}
    \hat{\tilde{R}}(\eta)=&\left\Vert \boldsymbol{p}_t-
    \boldsymbol{v}_0\eta_0+\frac{1}{2}\delta\boldsymbol{\dot{v}}_0\eta_0^{2}\right\Vert\\
    &+\frac{Q}
    {\left\Vert \boldsymbol{p}_t-\boldsymbol{v}_0\eta_0+\frac{1}{2}\delta\boldsymbol{\dot{v}}_0\eta_0^{2}\right\Vert}
    (\eta-\eta_0)^2\nonumber
\end{align}
where
\begin{align}
    Q =& 1.5(\delta\boldsymbol{\dot{v}}_0)^T\delta\boldsymbol{\dot{v}}_0\eta_0^2-
        3(\delta\boldsymbol{\dot{v}}_0)^T\boldsymbol{v}_0\eta_0\\
        &+\boldsymbol{p}_t^T\delta\boldsymbol{\dot{v}}_0+
        (\boldsymbol{v}_0)^T\boldsymbol{v}_0\nonumber
\end{align}

In the first term of the expansion, $\frac{1}{2}\delta\boldsymbol{\dot{v}}_0\eta_0^{2}$ causes a 
constant offset. For curve errors, this term causes a small shift in the imaged target. In practice,
this shift isn't strongly apparent, because the image degrades due to other terms before the shifting 
becomes strong. For phase errors, this constant offset doesn't affect the focus of the image.

In the second term of the expansion, $1.5(\delta\boldsymbol{\dot{v}}_0)^T\delta\boldsymbol{\dot{v}}_0$
is a quadratic error term and is considered very small. The 
$\frac{1}{2}\delta\boldsymbol{\dot{v}}_0\eta_0^{2}$ term in the denominator is small compared to other
terms in the denominator and causes negligible overall error. The 
$\boldsymbol{p}_t^T\delta\boldsymbol{\dot{v}}_0$ in the numerator causes a time-varying error. 
Interestingly, this error is in terms of the target location implying that the location of the target 
affects the severity of attitude imaging error. For both curve and phase errors, this term 
causes blurring similar to the along track velocity errors.

The $3(\delta\boldsymbol{\dot{v}}_0)^T\boldsymbol{v}_0\eta_0$ term also causes a time-varying
error. For both curve and phase errors, this again results in blurring; however, this term only becomes
significant for along track acceleration errors. This term is negligible for cross track and elevation
errors due to orthogonality. Note that this term is dependent on the time of closest approach and 
therefore changes depending on the location of the target in azimuth.

Notional illustrations of how attitude errors propagate through the SAR processing steps are shown in
Figures \ref{fig:roll_error} and \ref{fig:pitch_error}. It was shown that yaw errors do not affect
SAR images, therefore these figures depict only roll and pitch errors. Each figure is split into
subfigures with interpretations identical to those of the position and velocity error figures.

\section{Simulated Data}\label{sec:simulation}
The analysis presented in the previous section is now verified via simulation. SAR images are first
formed using the true trajectory. Initial errors are then injected and propagated to yield a corrupted
estimate of the trajectory. Images are formed with estimation errors and are compared to the truth 
reference image.  The presence and extent of shifting and blurring, as predicted by the development of
section \ref{sec:analysis}, is also verified.

\begin{figure}[!htb]
    \centering
    \includegraphics[height=2.5in]{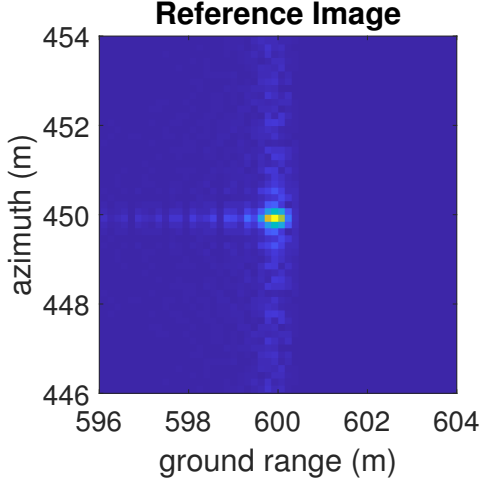}
    \caption{Reference image for simulated SAR data.}
    \label{fig:reference simulation}
\end{figure}

For each navigation error, a figure is presented with a simulated SAR image superimposed with the 
predicted target shift. The reference image to which each SAR chip should be compared is provided in
Figure \ref{fig:reference simulation}.

\begin{figure*}[p]
    \centering
    \includegraphics[width=2.1in]{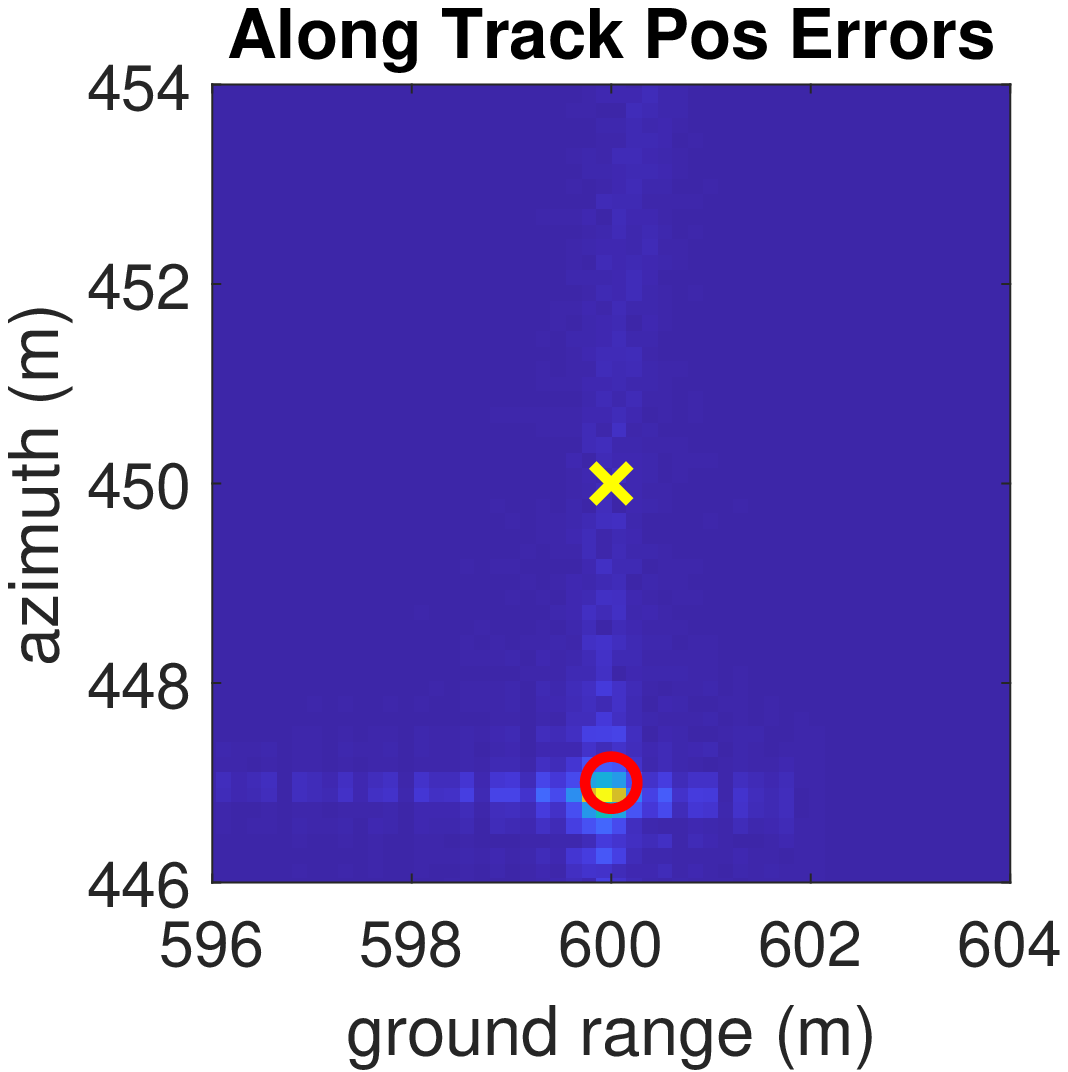}
    \includegraphics[width=2.1in]{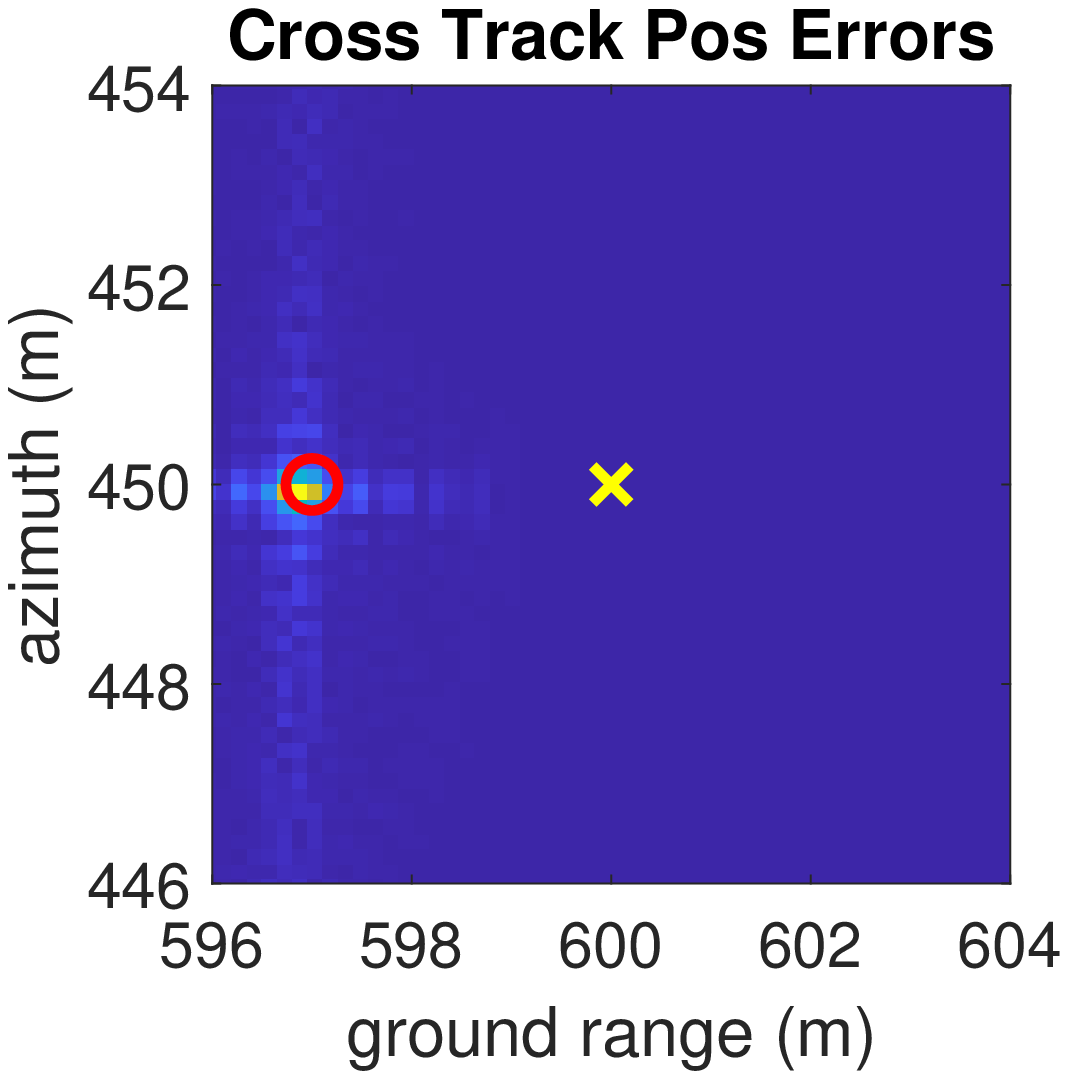}
    \includegraphics[width=2.1in]{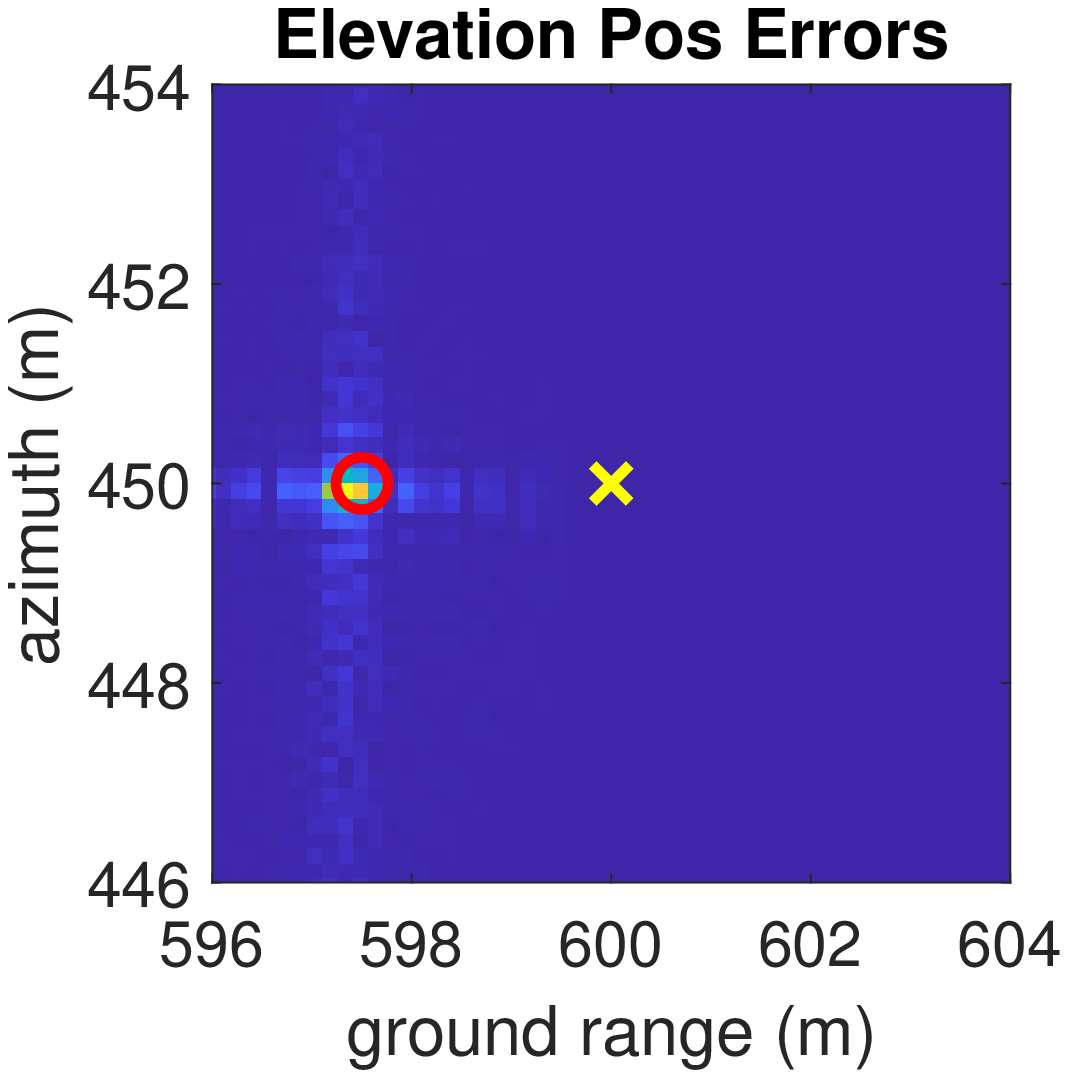}
    \caption{Position errors in simulated data: Left, along track position error (3 m). Middle, 
        cross track position error (3 m). Right, elevation position error (3 m). Each figure is superimposed with
        a reference target location, ``X'', and a predicted target shift, ``O''.}
    \label{fig:simulation pos}
\end{figure*}

\begin{figure*}[p]
    \centering
    \includegraphics[width=2.1in]{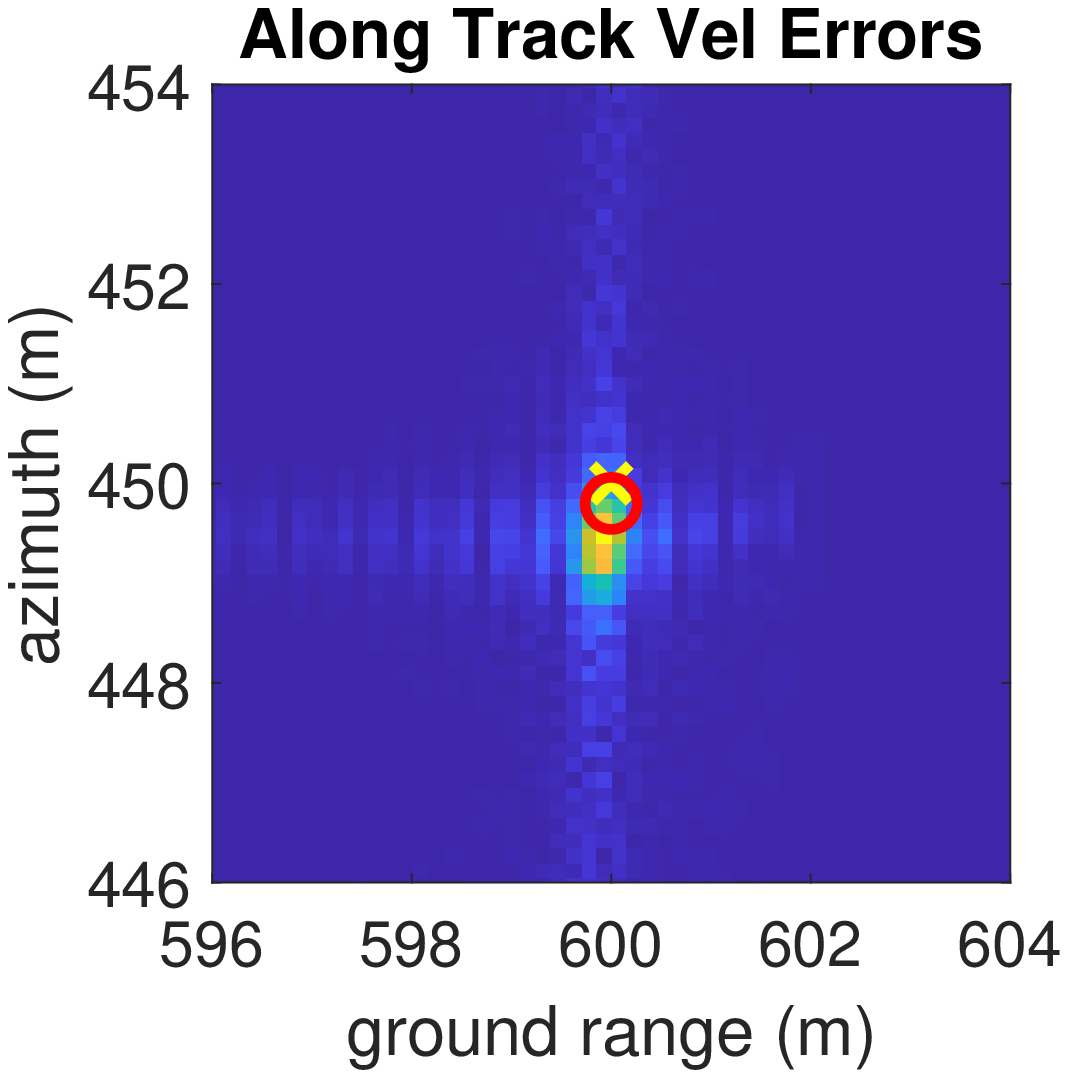}
    \includegraphics[width=2.1in]{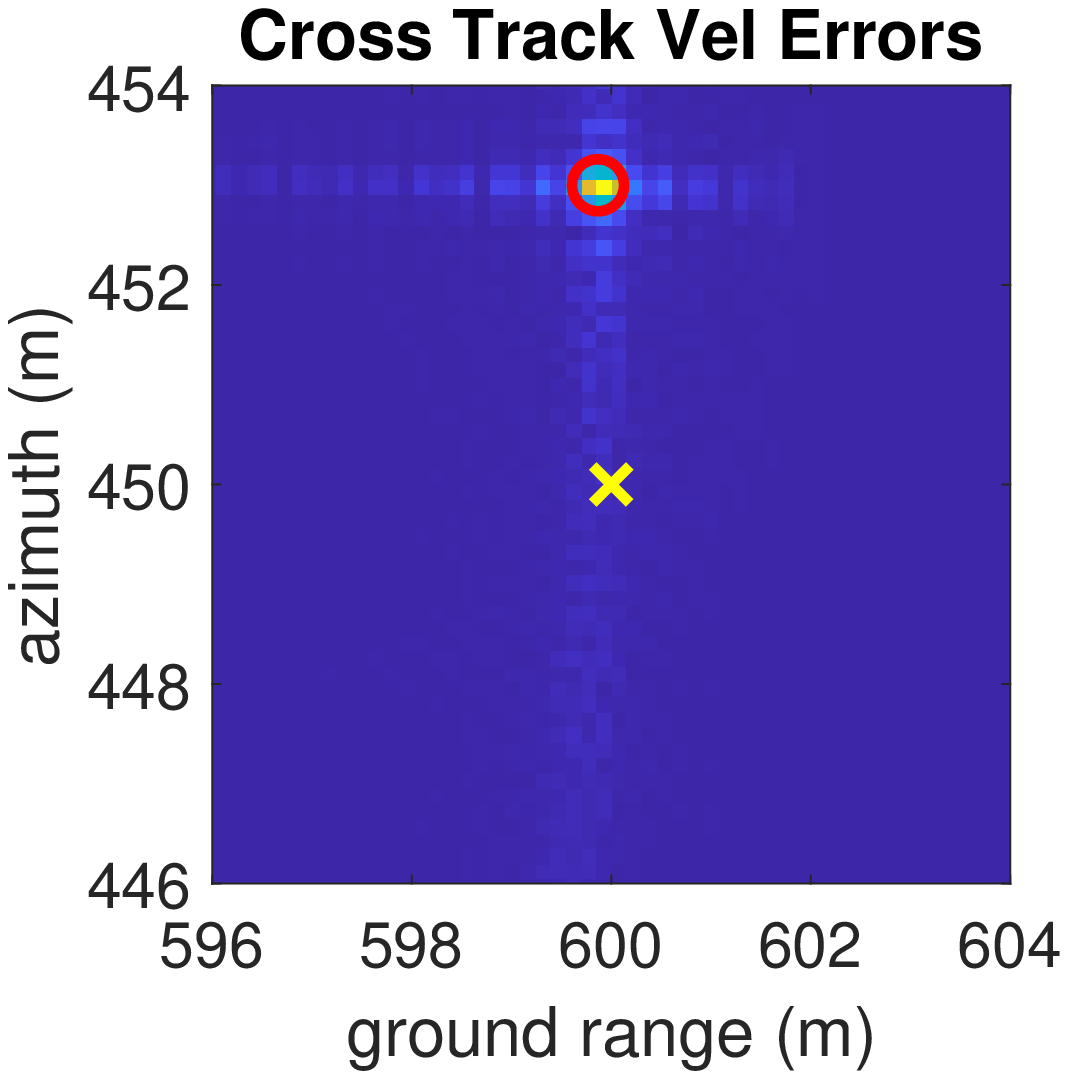}
    \includegraphics[width=2.1in]{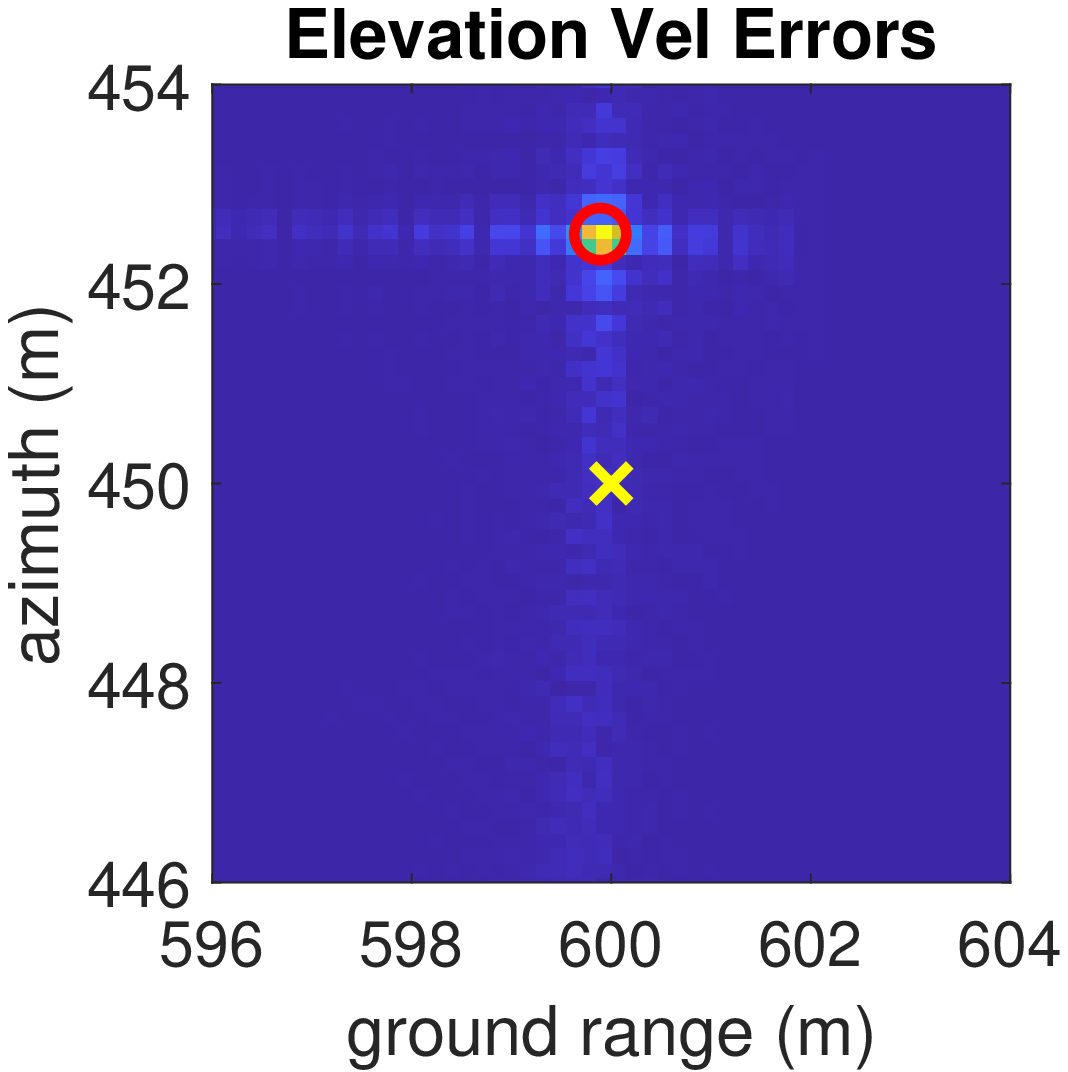}
    \caption{Velocity errors in simulated data: Left, along track velocity error (0.1 m/s). Middle, 
        cross track velocity error (0.05 m/s). Right, elevation velocity error (0.05 m/s). Each figure is superimposed with
        a reference target location, ``X'', and a predicted target shift, ``O''.}
    \label{fig:simulation vel}
\end{figure*}

\begin{figure*}[p]
    \centering
    \includegraphics[width=2.1in]{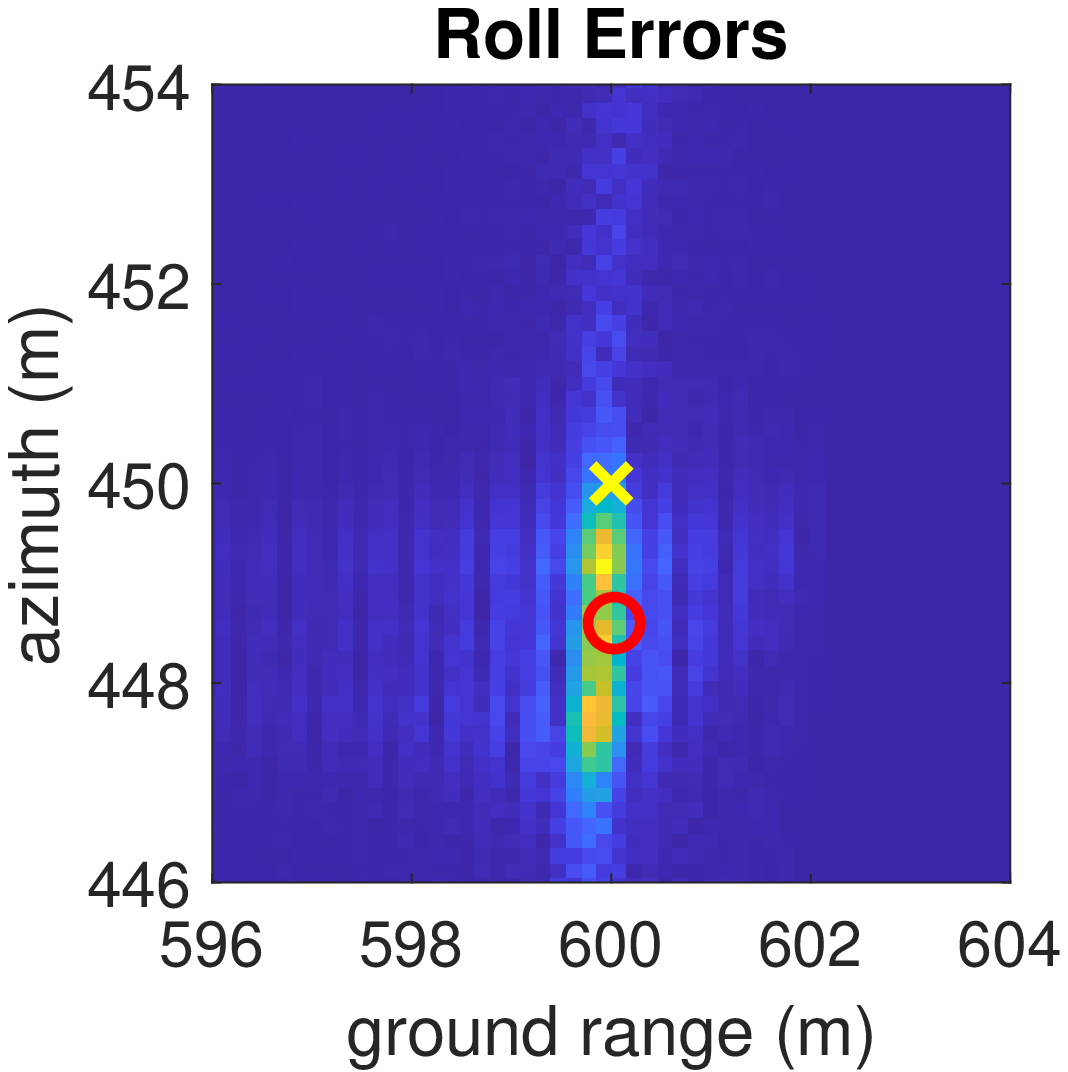}
    \includegraphics[width=2.1in]{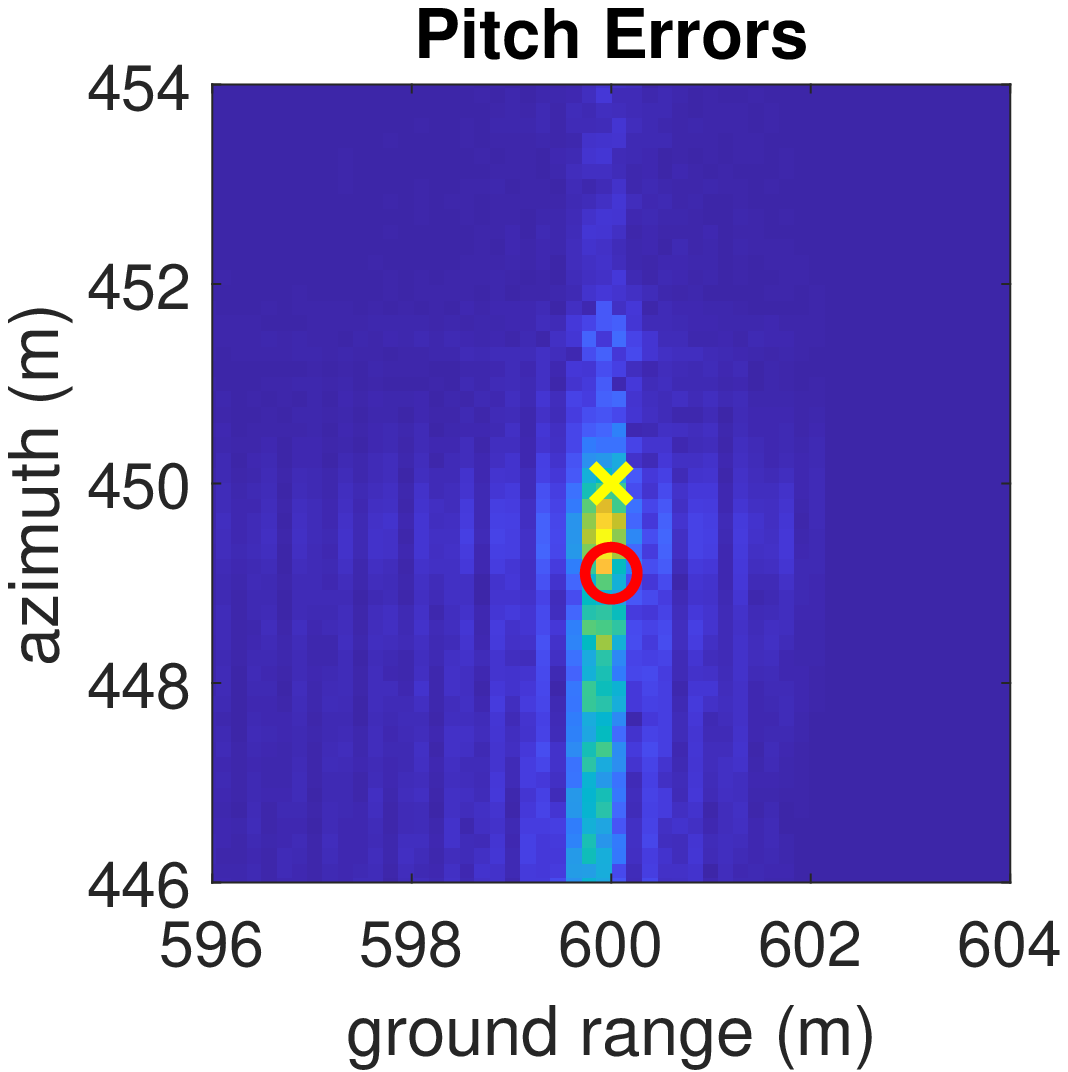}
    \includegraphics[width=2.1in]{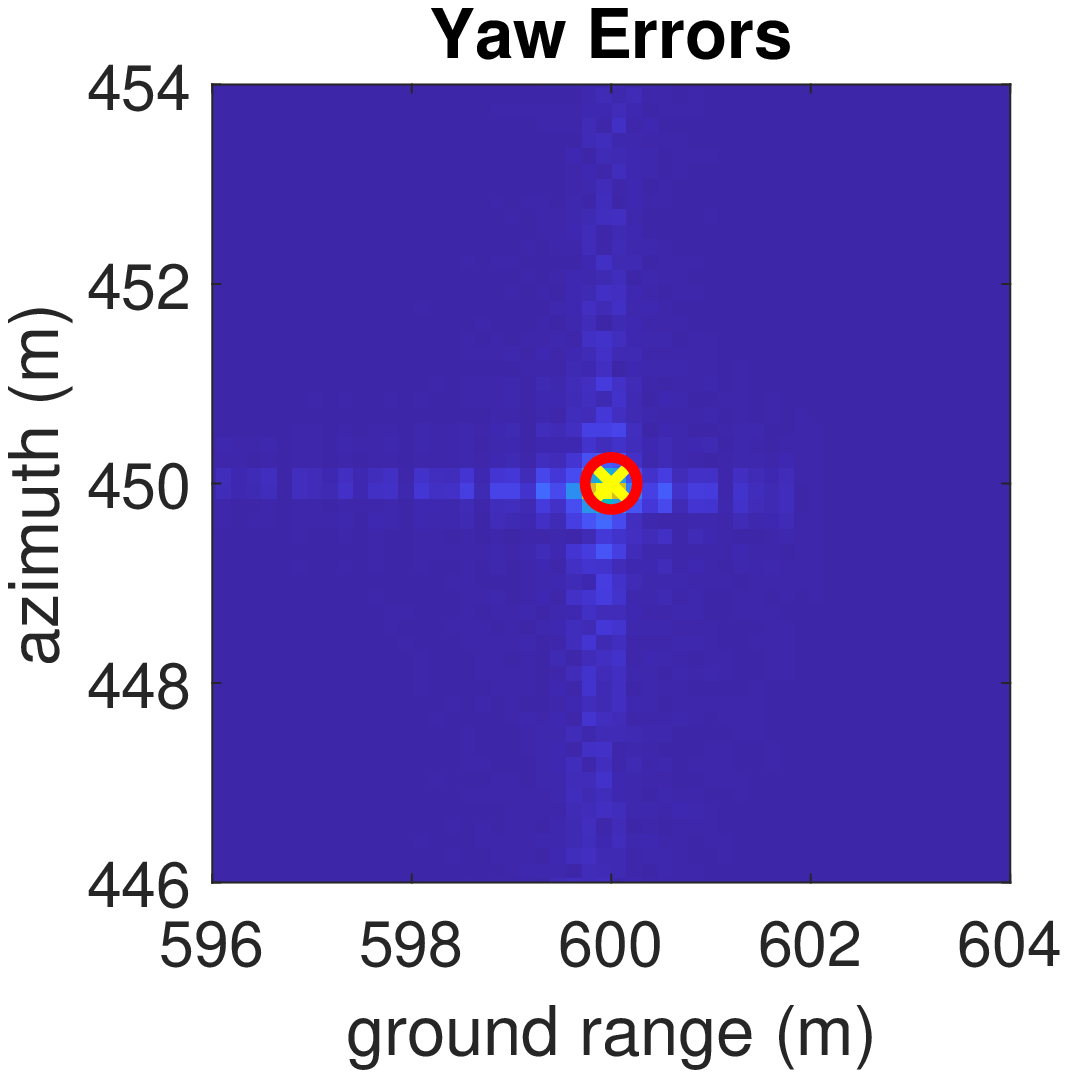}
    \caption{Attitude errors in simulated data: Left, roll error (0.001 rad). Middle, 
        pitch error (0.02 rad). Right, yaw error (0.1 rad). Each figure is superimposed with
        a reference target location, ``X'', and a predicted target shift, ``O''.}
    \label{fig:simulation att}
\end{figure*}

The SAR images formed given estimation errors are provided in Figures \ref{fig:simulation pos},
\ref{fig:simulation vel}, and \ref{fig:simulation att}.
For each image, a superimposed ``X'' shows the location of the reference target in relation to the current
image. A superimposed ``O'' shows the predicted location of the target given the injected estimation errors.

The ``X'' and ``O'' are generated using the true range equation, (\ref{eqn:time varying range}),
and the estimated range equation, (\ref{eqn:estimated range equation}), respectively. Equation (\ref{eqn:time varying range})
is used to find the true range of closest approach and time of closest approach, denoted $R_0$ and $\eta_0$.
Equation (\ref{eqn:estimated range equation}) is used to find the estimated range of closest 
approach and time of closest approach, denoted $\hat{R_0}$ and $\hat{\eta}_0$.
\begin{align}
    R_0=\min R(\eta),\quad & \hat{R}_0=\min \hat{R}(\eta)\\
    \eta_0 = \argmin_{\eta}R(\eta),\quad & \hat{\eta}_0 = \argmin_{\eta}\hat{R}(\eta)\nonumber
\end{align}
The range of closest approach and time of closest approach are used as coordinates 
to overlay ``X'' and ``O'' onto each image.

Figures \ref{fig:simulation pos}-\ref{fig:simulation att} illustrate that in all cases, the direction
of shifts and blurs is consistent with the development of section \ref{sec:analysis}.  Furthermore, 
in cases where blur is negligible, the amount of shift is accurately predicted utilizing the method 
described previously. It is important to highlight the ambiguity that exists in relating the SAR 
image error with the attributing navigation error.  From a single image, for example, it is impossible 
isolate the effects of cross track position and elevation errors, since both cause shifts in the cross
track position of the target.  Similar difficulties existing in attributing along-track shifts and 
blurs to the corresponding navigation errors.

\section{Real Data}\label{sec:real data}
\begin{figure}[!htb]
    \centering
    \includegraphics[height=2.5in]{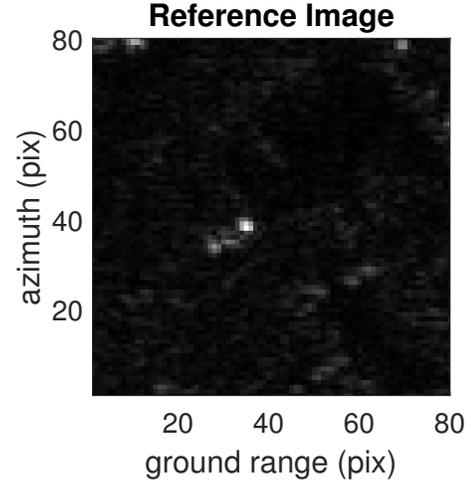}
    \caption{Reference image for real SAR data.}
    \label{fig:reference real}
\end{figure}

\begin{figure*}[p]
    \centering
    \includegraphics[width=2.1in]{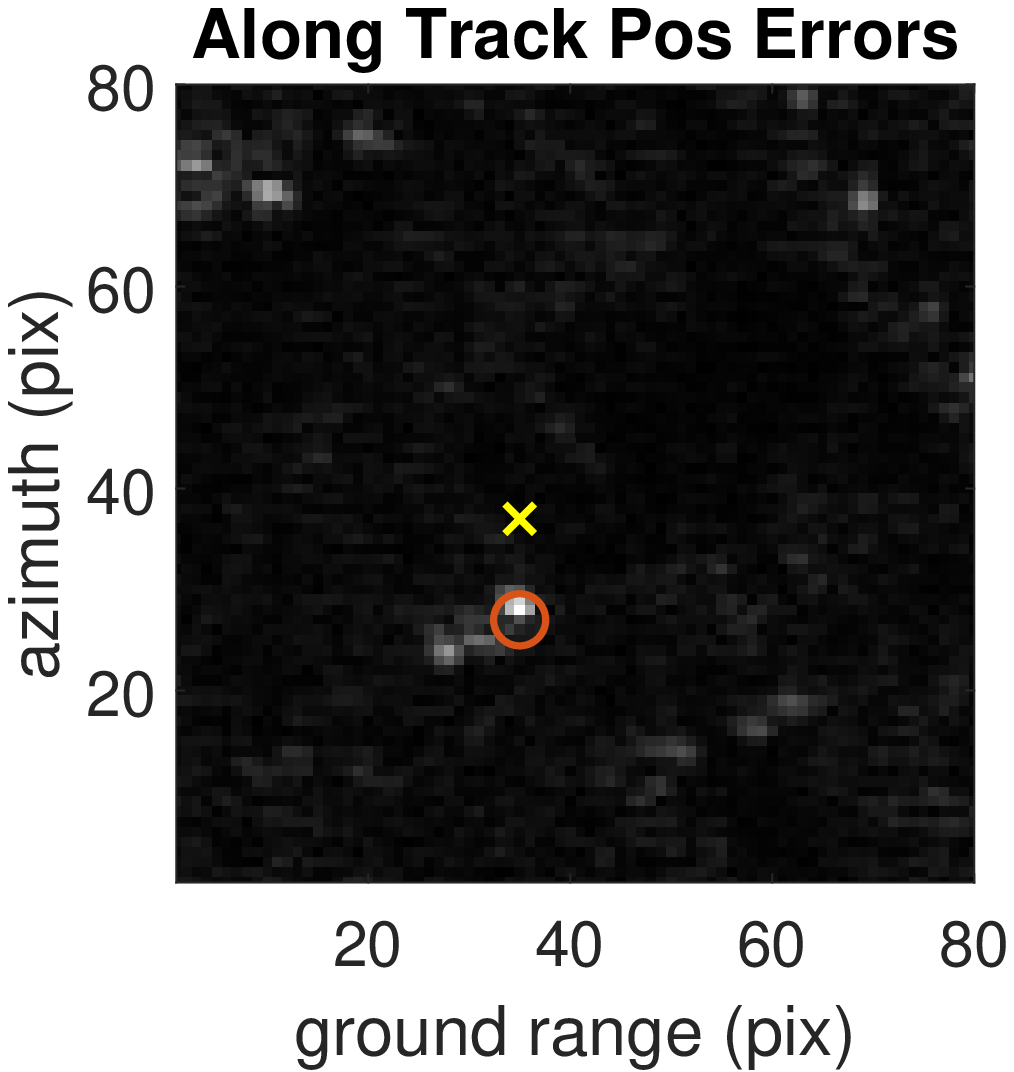}
    \includegraphics[width=2.1in]{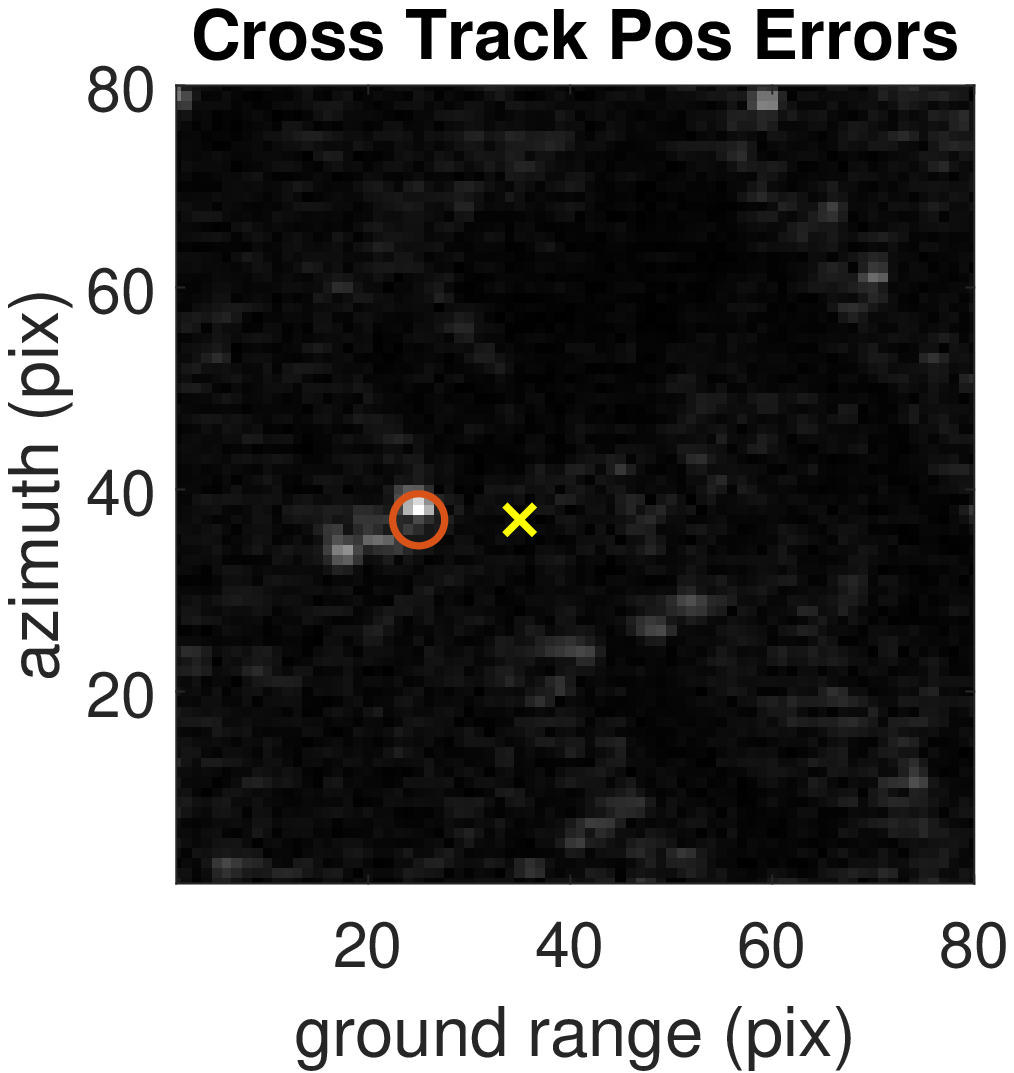}
    \includegraphics[width=2.1in]{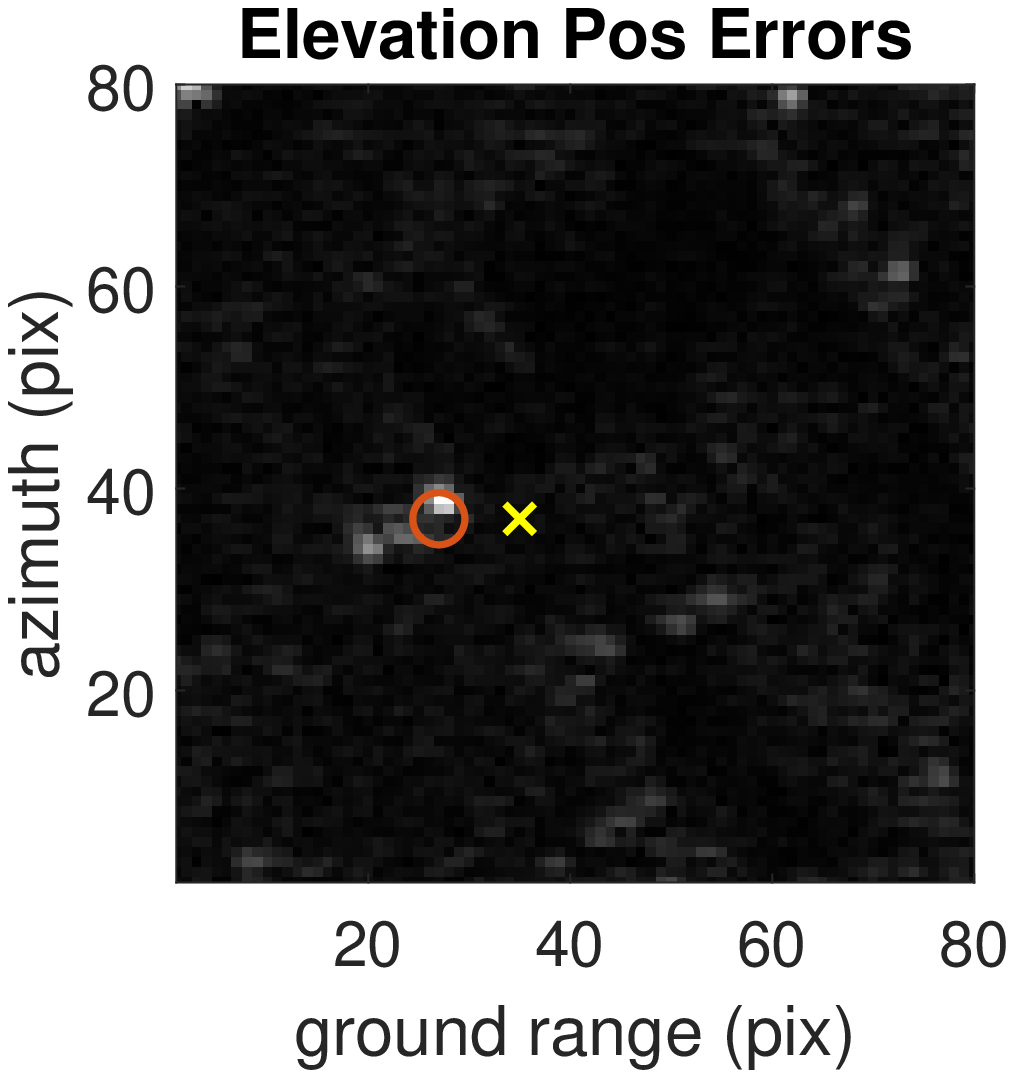}
    \caption{Position errors in real data: Left, along track position error (3 m). Middle, 
        cross track position error (3 m). Right, elevation position error (3 m). Each figure is superimposed with
        a reference target location, ``X'', and a predicted target shift, ``O''.}
    \label{fig:real pos}
\end{figure*}

\begin{figure*}[p]
    \centering
    \includegraphics[width=2.1in]{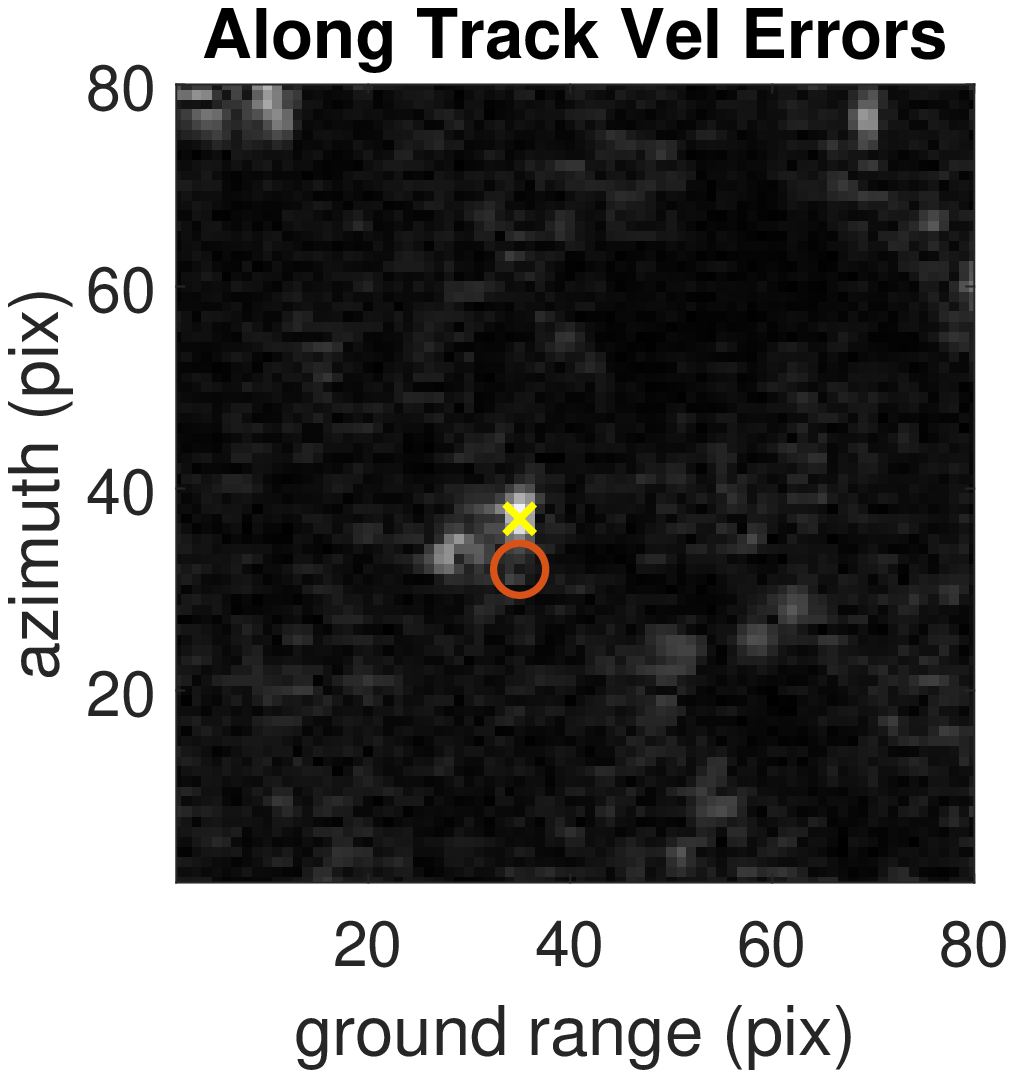}
    \includegraphics[width=2.1in]{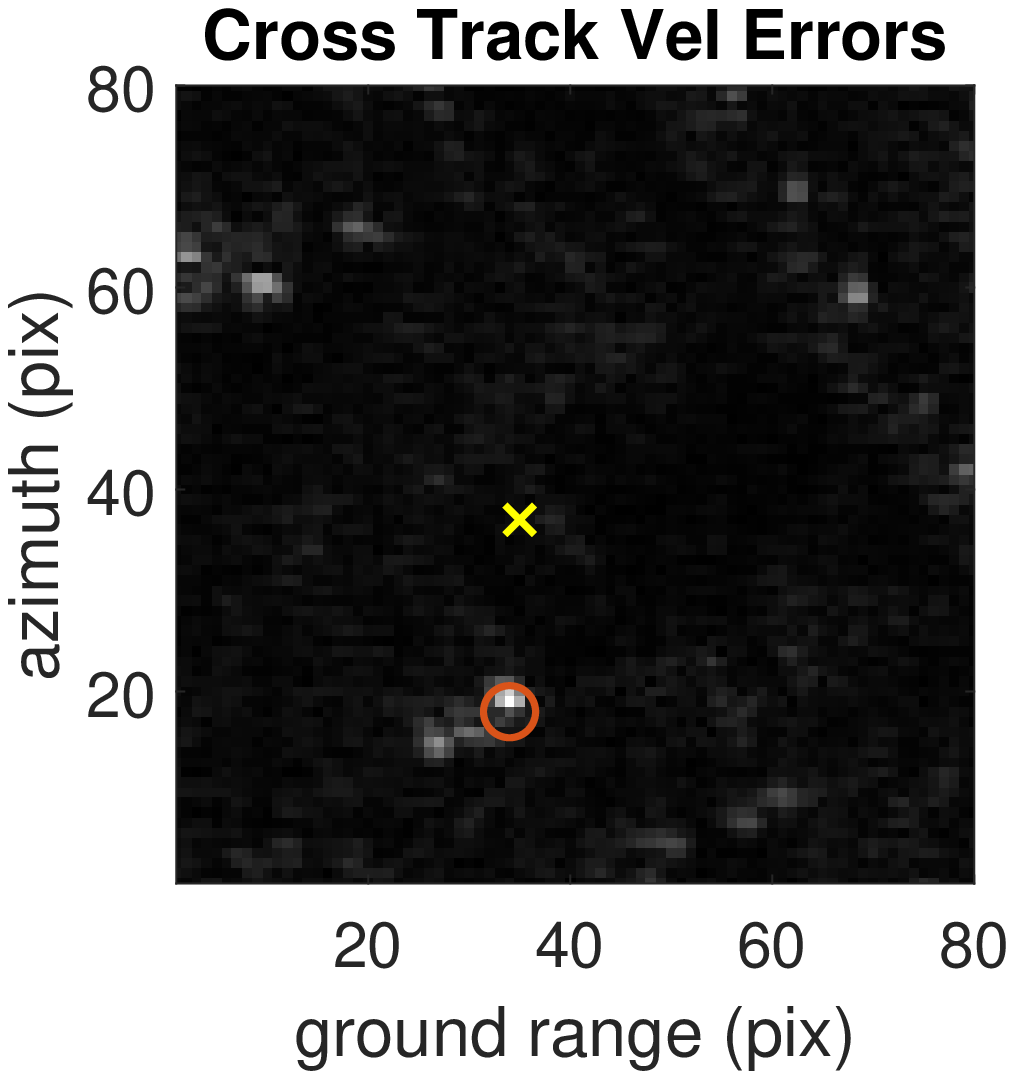}
    \includegraphics[width=2.1in]{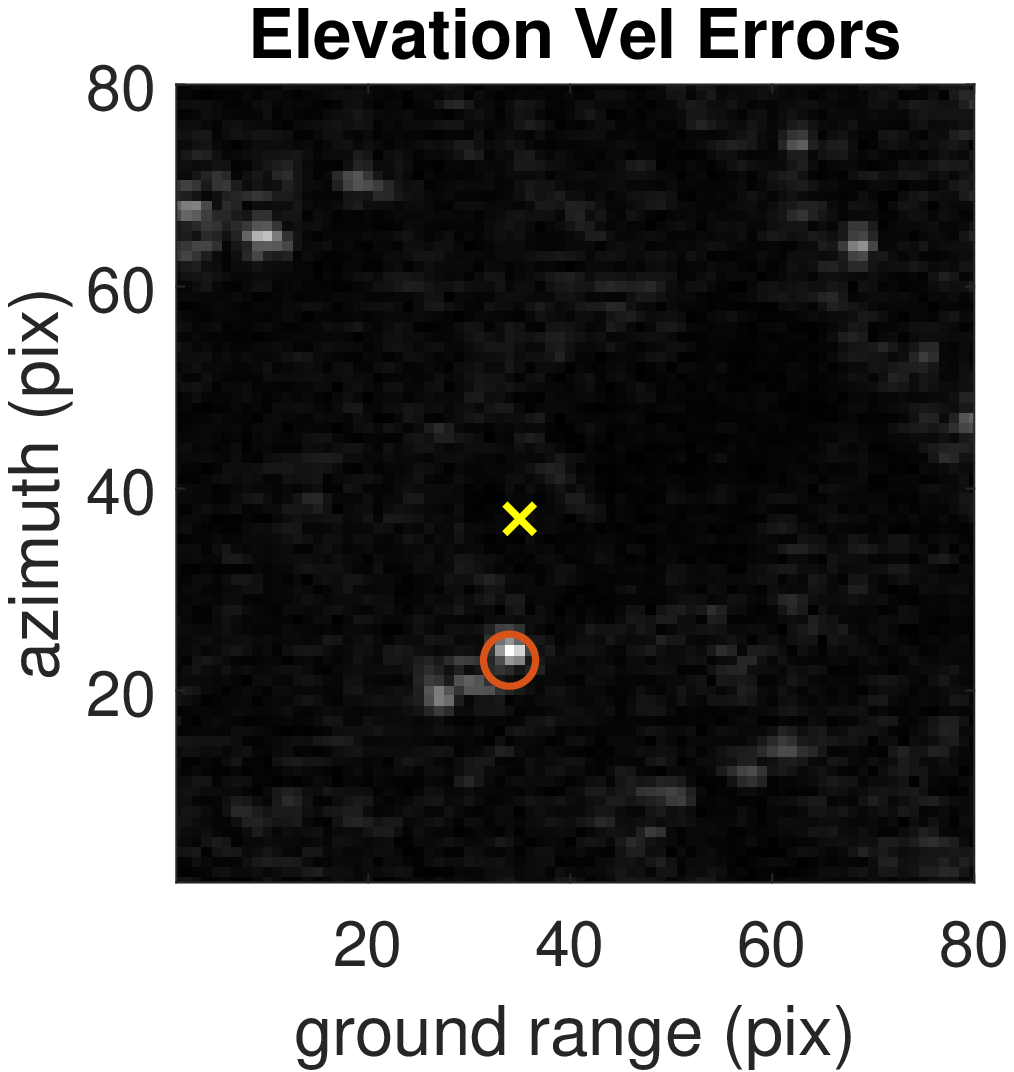}
    \caption{Velocity errors in real data: Left, along track velocity error (1 m/s). Middle, 
        cross track velocity error (0.2 m/s). Right, elevation velocity error (0.2 m/s). Each figure is superimposed with
        a reference target location, ``X'', and a predicted target shift, ``O''.}
    \label{fig:real vel}
\end{figure*}

\begin{figure*}[p]
    \centering
    \includegraphics[width=2.1in]{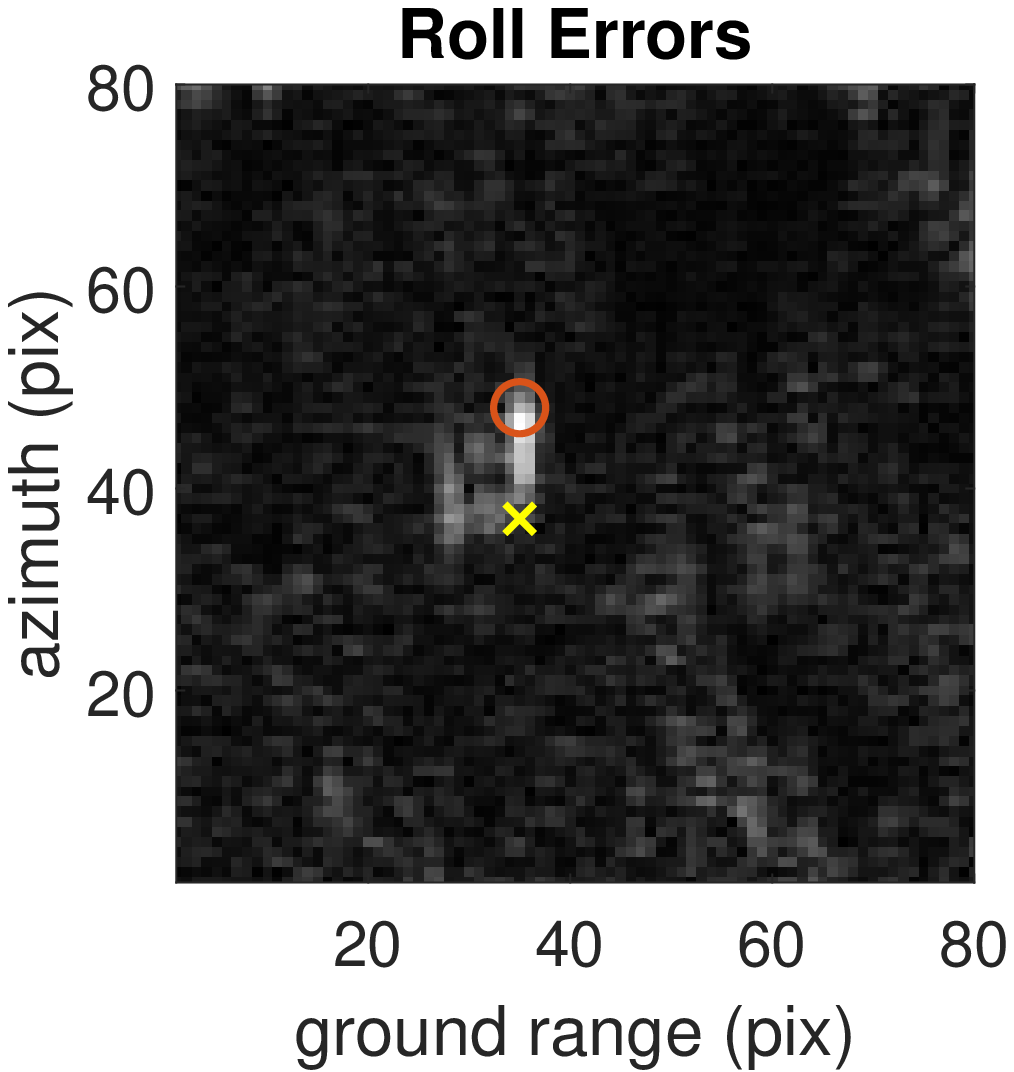}
    \includegraphics[width=2.1in]{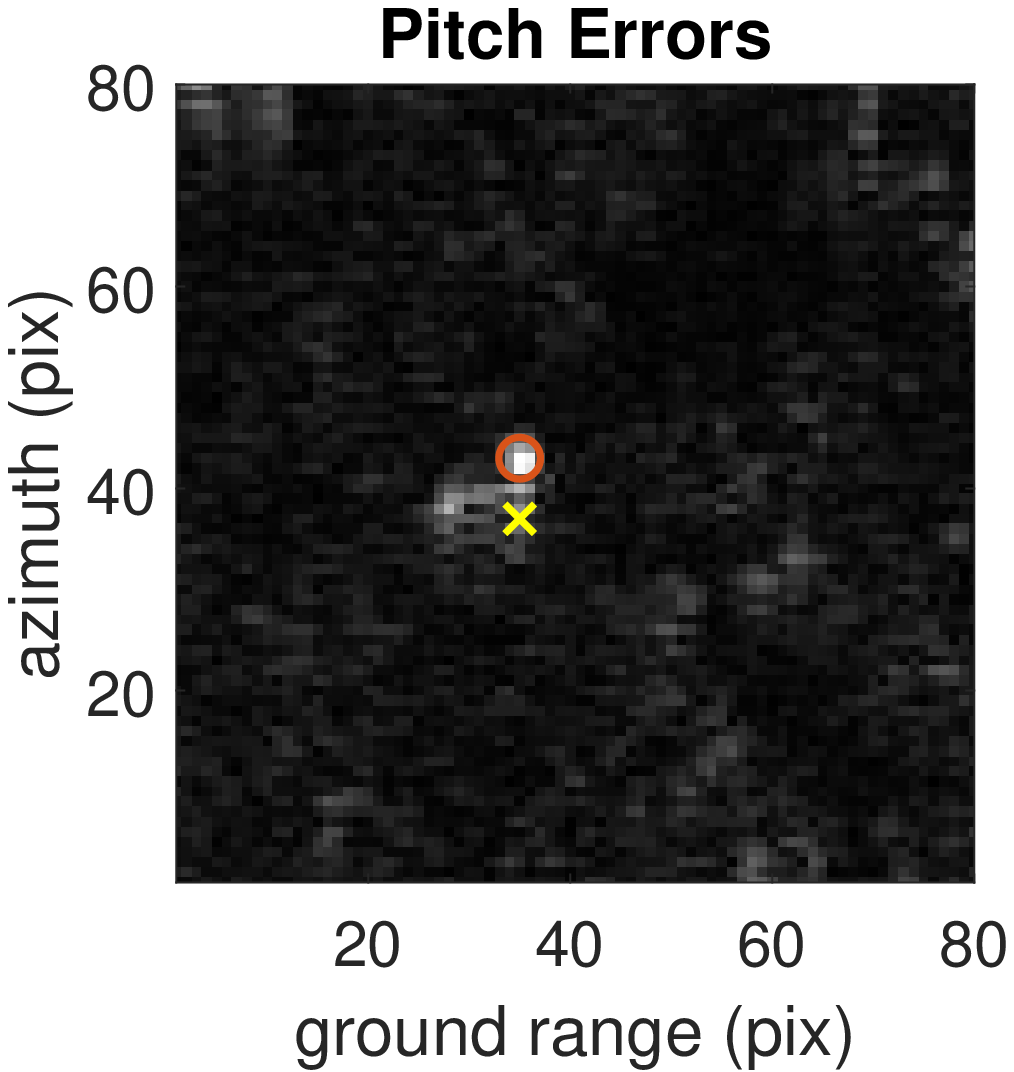}
    \includegraphics[width=2.1in]{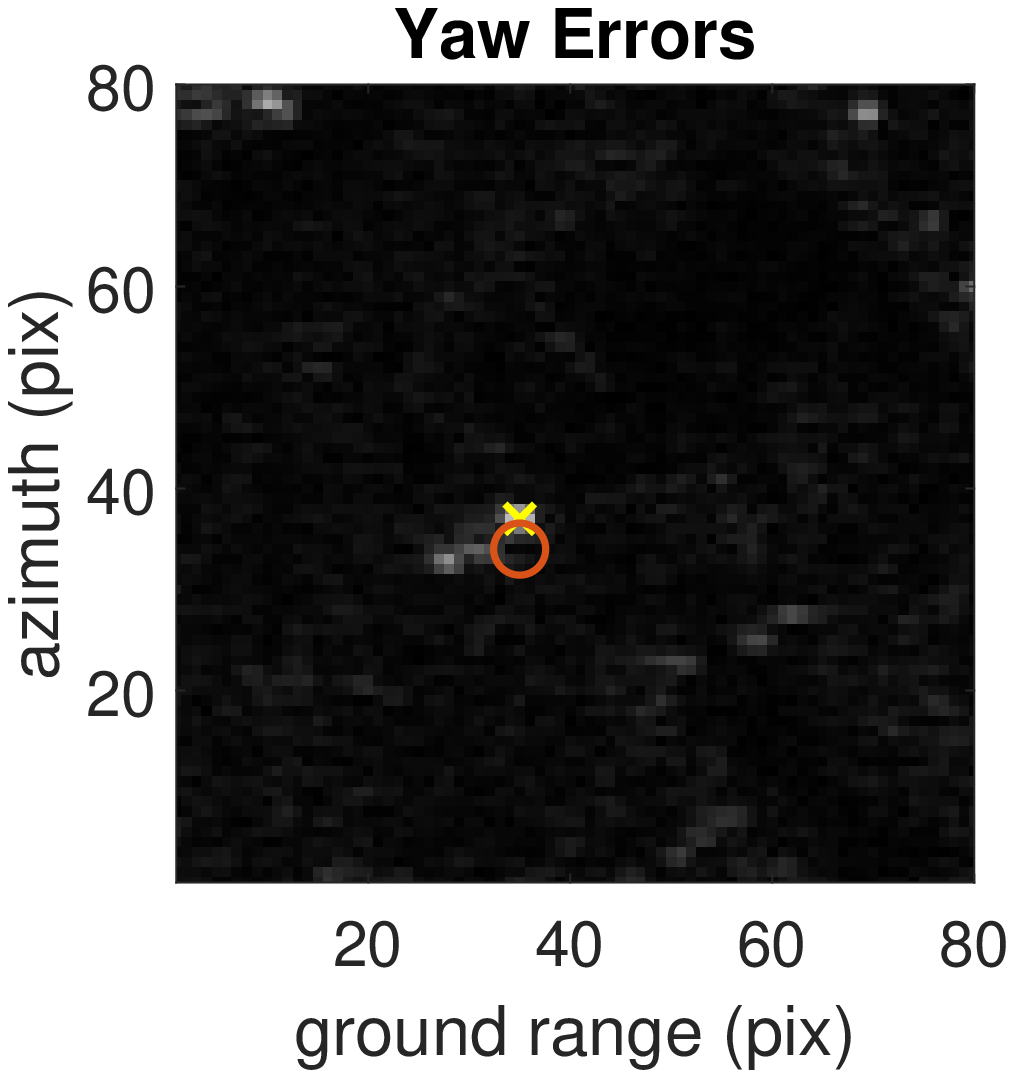}
    \caption{Attitude errors in real data: Left, roll error (0.01 rad). Middle, 
        pitch error (0.5 rad). Right, yaw error (0.1 rad). Each figure is superimposed with
        a reference target location, ``X'', and a predicted target shift, ``O''.}
    \label{fig:real att}
\end{figure*}

The analysis from Section \ref{sec:analysis} is be further verified using real SAR data. Radar data 
was collected in Logan, Utah. SAR images are formed using a post-processed, high fidelity navigation
solution, which is considered truth for the purposes of this research. The reference image in Figure 
\ref{fig:reference real} is formed using the truth trajectory. Each type of navigation error is 
then injected into the truth trajectory, from which the distorted SAR images of Figures 
\ref{fig:real pos}-\ref{fig:real att} are formed.

The results on real SAR data mirror the trends observed in section \ref{sec:simulation}. The type and 
direction of the shifts and blurs are consistent with the predictions of section \ref{sec:analysis}.  
In the case of negligible blurs, the shifts on real data are accurately predicted using the method 
described in section \ref{sec:simulation} for all cases except yaw error, where a very small 
prediction error is observed.  Finally, ambiguity in the attribution of error sources to image errors
is observed in the real SAR data.   Despite the small discrepancy in yaw, these results serve to 
further validate the relationships developed in section \ref{sec:analysis}.

\section{Conclusion}\label{sec:conclusion}
This paper analyzes errors in the formation of SAR images using the Back-Projection Algorithm from a
navigation perspective, for the case of straight-and-level flight. Relationships are developed between 
the position, velocity, and attitude estimation errors at the beginning of the synthetic aperture and 
the observed shifts and blurs of the corrupted BPA SAR image.

The developed relationships were observed and validated on both simulated and real SAR data.  In the 
case of negligible blurring, the location of the target in the corrupted SAR image is accurately 
predicted given knowledge of the attributing navigation error.  These results suggest that errors in 
BPA SAR images could potentially be used in reverse; i.e. image errors could be characterized and 
exploited as a navigation aid in GPS-denied applications.  For a single image, however, it was 
observed that the shifts/blurs are not unique to an individual navigation error. The presence of the 
target location in the developed relationships suggests that the effect of navigation errors can be 
modified by the selection of the target location.  One obvious extension of this work is the 
consideration of multiple targets with large geometric diversity, to resolve the ambiguity in 
attributing error sources. Furthermore, methods which characterize the amount and direction of image 
blurring must be developed to exploit the information contained therein.



\bibliography{SourcesErrorAnalysis}

\begin{thebibliography}{10}

\bibitem{cumming_wong}
I.~G. Cumming and F.~H. Wong, {\em Digital Processing of Synthetic Aperture
  Radar Data}.

\bibitem{duersch_analysis_2015}
M.~I. Duersch and D.~G. Long, ``Analysis of time-domain back-projection for
  stripmap {SAR},'' {\em International Journal of Remote Sensing}, vol.~36,
  pp.~2010--2036, Apr. 2015.
\newblock Publisher: Taylor \& Francis \_eprint:
  https://doi.org/10.1080/01431161.2015.1030044.

\bibitem{balamurugan}
G.~{Balamurugan}, J.~{Valarmathi}, and V.~P.~S. {Naidu}, ``Survey on uav
  navigation in gps denied environments,'' in {\em 2016 International
  Conference on Signal Processing, Communication, Power and Embedded System
  (SCOPES)}, pp.~198--204, 2016.

\bibitem{christensen_radar-aided_nodate}
R.~Christensen, J.~Gunther, and D.~Long, ``Radar-{Aided}, {GPS}-{Denied}
  {Navigation}.''

\bibitem{bamler_comparison_1992}
R.~Bamler, ``A comparison of range-{Doppler} and wavenumber domain {SAR}
  focusing algorithms,'' {\em IEEE Transactions on Geoscience and Remote
  Sensing}, vol.~30, pp.~706--713, July 1992.

\bibitem{farrell_effects_1973}
J.~L. Farrell, J.~H. Mims, and A.~Sorrell, ``Effects of {Navigation} {Errors}
  in {Maneuvering} {SAR},'' {\em IEEE Transactions on Aerospace and Electronic
  Systems}, vol.~AES-9, pp.~758--776, Sept. 1973.

\bibitem{chen_time-frequency_1998}
V.~C. Chen, ``Time-frequency analysis of {SAR} images with ground moving
  targets,'' in {\em Wavelet {Applications} {V}}, vol.~3391, pp.~295--302,
  International Society for Optics and Photonics, Mar. 1998.

\bibitem{moreira_estimating_1990}
J.~Moreira, ``Estimating the residual error of the reflectivity displacement
  method for aircraft motion error extraction from {SAR} raw data,'' in {\em
  {IEEE} {International} {Conference} on {Radar}}, pp.~70--75, May 1990.

\bibitem{xing_motion_2009}
M.~Xing, X.~Jiang, R.~Wu, F.~Zhou, and Z.~Bao, ``Motion {Compensation} for
  {UAV} {SAR} {Based} on {Raw} {Radar} {Data},'' {\em IEEE Transactions on
  Geoscience and Remote Sensing}, vol.~47, pp.~2870--2883, Aug. 2009.
\newblock Conference Name: IEEE Transactions on Geoscience and Remote Sensing.

\bibitem{bing_image_2013}
S.~Bing, W.~Ye, C.~Jie, W.~Yan, and L.~Bing, ``Image position analysis of
  motion errors for missile-borne {SAR} based on diving model,'' in {\em 2013
  {IEEE} {International} {Conference} on {Imaging} {Systems} and {Techniques}
  ({IST})}, pp.~206--209, Oct. 2013.
\newblock ISSN: 1558-2809.

\bibitem{grewal_global_2013}
M.~S. Grewal, A.~P. Andrews, and C.~Bartone, {\em Global navigation satellite
  systems, inertial navigation, and integration}.
\newblock Hoboken: John Wiley \& Sons, third edition~ed., 2013.

\bibitem{farrell_aided_2008}
J.~Farrell, {\em Aided navigation: {GPS} with high rate sensors}.
\newblock Electronic engineering, New York: McGraw-Hill, 2008.
\newblock OCLC: ocn212908814.

\bibitem{savage_strapdown_2000}
P.~G. Savage, {\em Strapdown analytics}.
\newblock Maple Plain, Minn.: Strapdown Associates, 2000.

\bibitem{zanetti_rotations_2019}
R.~Zanetti, ``Rotations, {Transformations}, {Left} {Quaternions}, {Right}
  {Quaternions}?,'' {\em The Journal of the Astronautical Sciences}, vol.~66,
  pp.~361--381, Sept. 2019.

\bibitem{maybeck_stochastic_1994}
P.~S. Maybeck, {\em Stochastic {Models}, {Estimation}, and {Control}}, vol.~1.
\newblock New York: Navtech Book and Software Store, 1994.

\end{thebibliography}
\bibliographystyle{ieeetr}
\end{document}